\documentclass{article}
\usepackage[preprint]{neurips_2024}

\usepackage{fancyhdr} 
\usepackage{blindtext} 
\usepackage{makecell}
\usepackage{natbib}

\usepackage{fontawesome5} 

\usepackage[utf8]{inputenc} 
\usepackage[T1]{fontenc}    
\usepackage{hyperref}       
\usepackage{url}            
\usepackage{booktabs}       
\usepackage{amsfonts}       
\usepackage{nicefrac}       
\usepackage{wrapfig}
\usepackage{xcolor}         
\usepackage{amsmath}
\usepackage{float}
\usepackage{tabularray}
\usepackage{multirow}
\usepackage{booktabs, caption, array}
\usepackage{amssymb}
\usepackage{lipsum}         
\usepackage{CJKutf8}        
\usepackage{nomencl}
\usepackage{glossaries}
\usepackage{xspace}
\usepackage{multirow}
\usepackage{graphicx}
\usepackage{bbding}
\usepackage{pifont}
\usepackage{algorithm}
\usepackage{enumitem}
\usepackage{multicol}
\usepackage{caption}
\usepackage{array}
\usepackage{fp}
\usepackage{multirow}   
\usepackage{makecell}
\usepackage{flushend}
\usepackage{cases}
\usepackage{color}
\usepackage{algpseudocode}
\usepackage{listings}
\usepackage{mathrsfs}
\usepackage{longtable}
\usepackage{colortbl}
\usepackage{bm}
\usepackage{caption}
\usepackage{natbib}
\usepackage{tabularx}

\usepackage{subcaption} 
\setlist[itemize]{noitemsep, topsep=0pt}
\usepackage{arydshln}
\usepackage{lipsum}
\definecolor{codegreen}{rgb}{0,0.3,0.6}
\definecolor{codegray}{rgb}{0.5,0.5,0.5}

\newcommand{\ignore}[1]{}

\usepackage[most]{tcolorbox} 
\usepackage{microtype}      
\definecolor{darkorange}{RGB}{255, 140, 0}
\definecolor{lightgreen}{RGB}{145, 204, 117}
\definecolor{lightyellow}{RGB}{250, 200, 88}
\definecolor{lightred}{RGB}{238, 102, 102}
\definecolor{lightblue}{RGB}{115, 192, 222}
\newtcolorbox{promptbox}[3][Judge Prompt]{
colback=black!5!white,
arc=5pt, 
boxrule=0.5pt,
fonttitle=\bfseries,
title=#1, 
before upper={\small}, fontupper=\fontfamily{ptm}\selectfont,
colframe=#2,
label=#3,
}
\definecolor{gray_1}{HTML}{B7B7B7}
\definecolor{gray_2}{HTML}{F0F0F0} 
\definecolor{frame_blue}{HTML}{A9D18E}

\newtcolorbox[auto counter, number within=section]{PromptBoxNew}[2][]{
    enhanced,
    breakable,
    colback=gray_2, 
    colframe=gray_1,
    coltitle=white,
    fontupper=\small,
    fonttitle=\bfseries,
    title={#2}, 
    label={#1},
    arc=2pt,
    boxrule=1pt,
    left=2mm, right=2mm, top=2mm, bottom=2mm,
}

\newtcolorbox[auto counter, number within=section]{PromptBox}[2][]{
    enhanced,
    breakable,
    colback=gray_2, 
    colframe=gray_1,
    coltitle=white,
    fontupper=\small,
    fonttitle=\bfseries,
    title={#2}, 
    label={#1},
    arc=2pt,
    boxrule=1pt,
    left=2mm, right=2mm, top=2mm, bottom=2mm,
}

\newtcolorbox{tracebox}[1]{
  enhanced,
  width=\linewidth,
  colback=white,
  colframe=black,
  coltitle=white,
  boxrule=0.9pt,
  arc=2pt,
  left=4mm,
  right=4mm,
  top=5.5mm,
  bottom=3mm,
  title={#1},
  fonttitle=\normalsize,
  attach boxed title to top left={xshift=3mm,yshift=-2.2mm},
  boxed title style={
    colback=black,
    colframe=black,
    boxrule=0pt,
    arc=2pt,
    left=2.5mm,
    right=2.5mm,
    top=1mm,
    bottom=1mm
  }
}

\newcommand{\tracefield}[2]{%
  \makebox[8.5em][l]{\quad #1:}#2\par
}

\usepackage{etoolbox}
\makeatletter
\patchcmd{\@maketitle}
  {\@toptitlebar}
  {%
    \hbox to \textwidth{%
    
    \includegraphics[height=0.75cm]{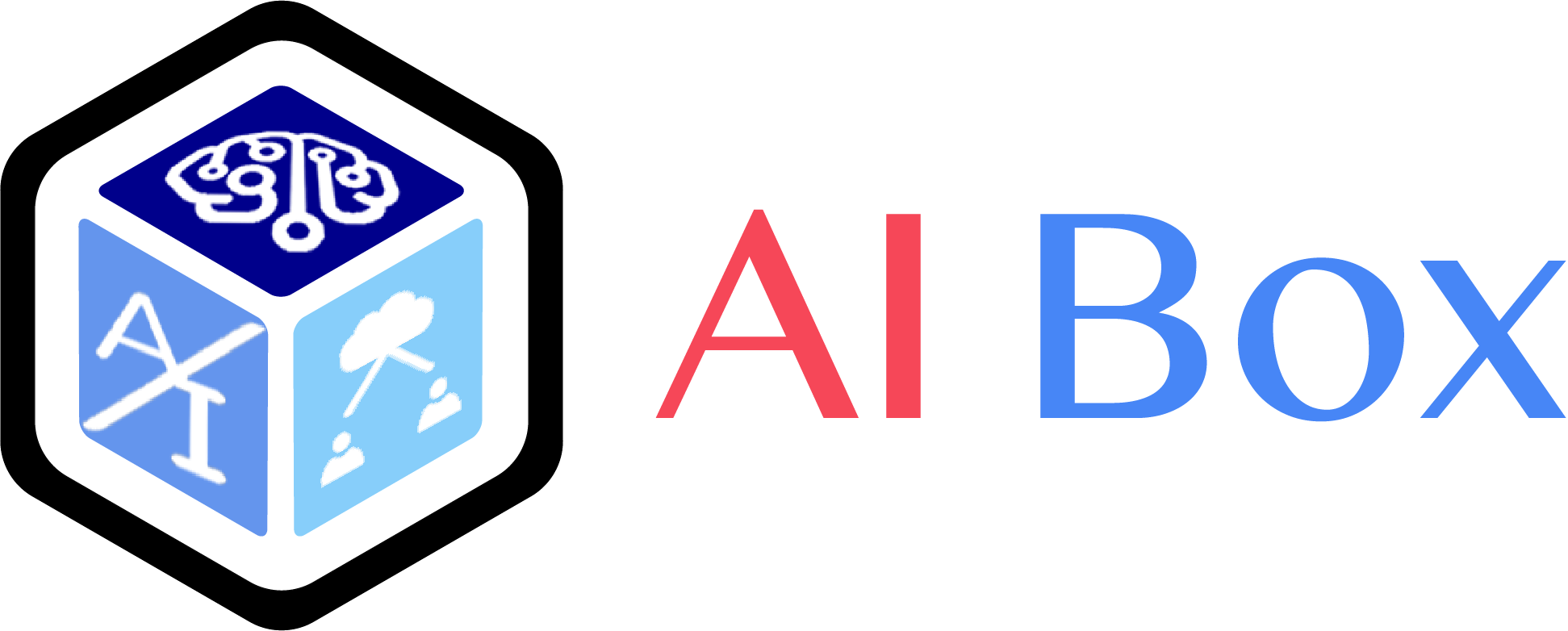}%
    \hfill
    \includegraphics[height=0.7cm]{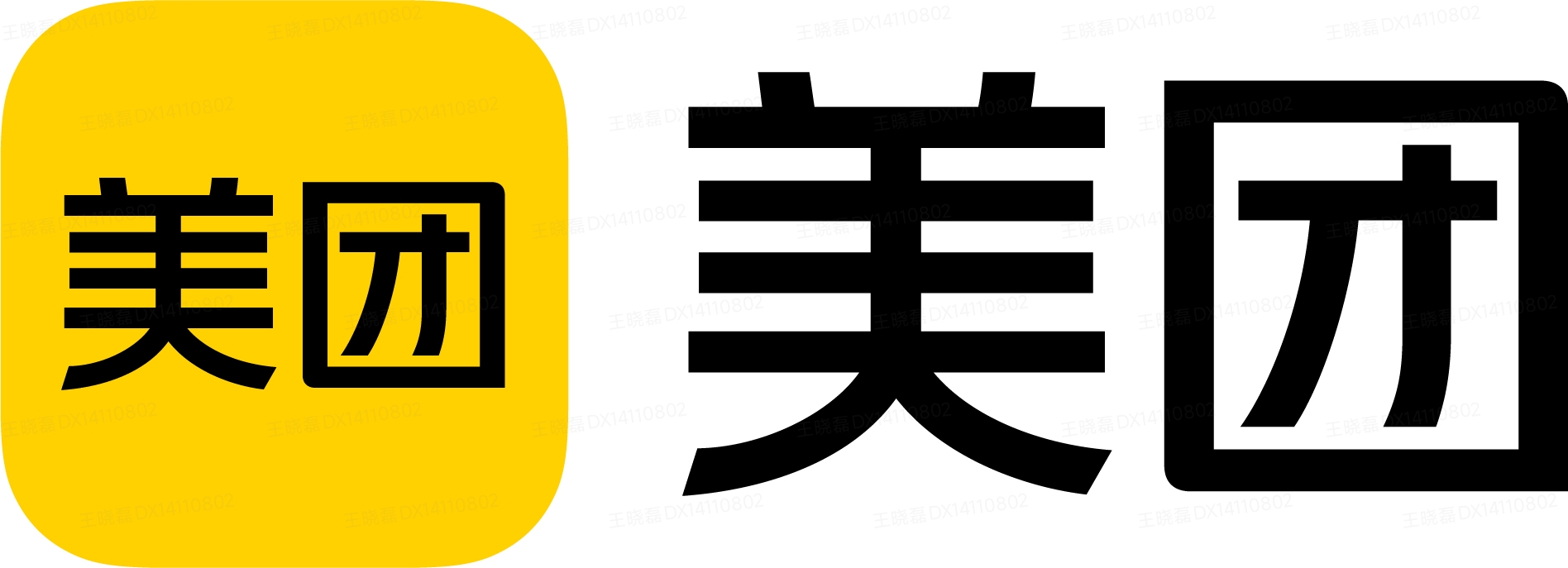}%
    }%
    \vspace{0.3em}
    \@toptitlebar
  }
  {}{}
\makeatother

\title{ClawRec: A Claw-Native Recommender System}

\author{%
  Chenghao Wu$^{1}$\thanks{Equal contribution.}~, 
  Kesha Ou$^{1*}$, 
  Xiaolei Wang$^{1*}$, 
  Bowen Zheng$^{1*}$,
  Bingqian Li$^{1}$,
  Enze Liu$^{1}$, \\
  \textbf{Wayne Xin Zhao$^{1}$\thanks{Corresponding author.}~,} 
  \textbf{Weitao Li$^{2}$,}
  \textbf{Long Zhang$^{2}$,}
  \textbf{Sheng Chen$^{2}$,}
    \textbf{Ji-Rong Wen$^{1}$}
  \\
  $^1$Gaoling School of Artificial Intelligence, Renmin University of China, $^2$Meituan.\\
  \texttt{wuchenghao@ruc.edu.cn},
\texttt{batmanfly@gmail.com}
}

\begin{document}
\maketitle

\definecolor{babyblue}{HTML}{F8F9FE}
\newtcolorbox{bluebox}{
  colback=babyblue,    
  colframe=babyblue,  
  width=1.0\textwidth,  
  center,               
  arc=8pt,                 
  boxrule=0pt,           
  boxsep=0pt,       
  left=2pt,               
  right=2pt,              
  top=10pt,                
  bottom=10pt              
}


\begin{bluebox}
\begin{abstract}
Recommender systems have become integral to navigating the modern digital ecosystem.
Yet most deployed systems remain confined within single-platform boundaries, observing localized interaction traces and ranking items from isolated candidate spaces.
This design is poorly suited to real-world tasks that unfold through searches, content consumption, and comparisons across multiple information sources.
Claw-style personal agents, with persistent access to authorized cross-platform context, create an opportunity for recommendation to operate around the user rather than any single platform.
In this paper, we introduce \emph{Claw-native recommender systems}, a new paradigm that moves beyond platform-local ranking to produce unified, complementary recommendation slates spanning diverse sources and content forms.
To instantiate this paradigm, we present \textbf{ClawRec}, the first recommender system designed to operate natively in this environment.
ClawRec maintains an evidence-linked, temporally structured user state that connects cross-platform behaviors with cross-source recommendations.
It organizes retrieval around functional source roles and selects candidates according to their marginal utility, producing non-redundant slates aligned with the user's active task.
To enable rigorous evaluation, we introduce \textbf{ClawRec-SimBench}, a benchmark constructed from sequences of concrete life events and cross-platform behavior trajectories.
Experiments show that ClawRec outperforms the strongest baselines, achieving an NDCG@20 of 0.6134 (+0.1126) and a Hit@20 of 0.6944 (+0.0854), while also improving user state quality and temporal alignment.
Our code and dataset are available at \url{https://github.com/RUCAIBox/ClawRec}.

\end{abstract}
\end{bluebox}

\section{Introduction}
\label{sec:introduction}

User needs often span multiple platforms, whereas most recommender systems operate within one.
Conventional recommendation is commonly formulated as in-platform item ranking: given a user's interaction history on a particular platform, the system ranks candidates from the corresponding item space~\cite{rendle2009bpr,kang2018selfattentive,sun2019bert4rec}.
This formulation is effective when a user's current need is both expressed and served within the same platform.
In practice, however, a user's current task may unfold through searches, content visits, comparisons, revisits, and feedback distributed across different information sources.
These actions form a coherent trajectory from the user's perspective but appear as fragmented and weakly connected traces to each individual platform.
As a result, a platform-local recommender observes only part of the current task and can respond only through its own candidate space.
This mismatch motivates a broader question: \emph{can recommender systems operate natively in the user's digital environment rather than around the partial view of a single platform?}

Existing research has taken important steps toward integrating fragmented user evidence.
Multi-behavior recommendation aggregates heterogeneous feedback, cross-domain methods transfer information across predefined domains, user lifecycle models capture longer-term evolution, and LLM-based or agentic methods incorporate natural-language context and multi-step reasoning~\cite{jin2020multibehavior,zang2022survey,li2023stan,wu2024survey,peng2025survey}.
Collectively, these directions enrich the evidence available for personalization.
However, they generally retain a predefined boundary around what the system can observe and what it can recommend: the relevant platforms, domains, data interfaces, and candidate spaces are specified by the application.
They therefore integrate fragmented evidence within an existing recommendation environment, rather than making the user's broader digital environment the operating scope of recommendation.

The recent emergence of personal agent frameworks such as OpenClaw~\cite{openclaw2026} creates a technical opportunity to move beyond this boundary.
We refer to such personal agents as \emph{Claw-style personal agents}.
These agents can persistently access authorized context and information sources distributed across the user's digital environment.
To realize this opportunity, we introduce the \emph{Claw-native recommender system}, which organizes recommendations around the user's current task rather than around any single platform.
Its input consists of the authorized evidence accessible to the agent, including cross-platform behaviors, relevant history, explicit chat context, and feedback on prior recommendations.
Its output is a unified ranked slate of complementary items drawn from multiple sources and content forms.
For example, the slate may include explanatory information, practical guidance, authoritative references, and product options.
These items are selected to provide complementary support for the user's current task.
From the user's perspective, this approach replaces repeated platform-by-platform search and manual synthesis with a single, task-centered recommendation experience whose candidate space is not confined to any one platform.

Realizing this paradigm introduces distinct technical challenges for recommender systems, driven by three key factors.
First, cross-platform interaction logs are temporally mixed and semantically heterogeneous.
A user's history combines transient searches, content visits, and completed tasks, making it difficult for standard sequence models or raw memory buffers to isolate the active task from persistent preferences and stale background noise.
Second, open-environment sources serve distinct functional roles, such as official verification, community experience, or product comparison, even when addressing the same topic.
Existing relevance-retrieval methods treat these sources as uniform item pools, failing to map specific task requirements to functionally appropriate information channels.
Third, supporting a real-world task requires a complementary recommendation slate rather than multiple variations of topically relevant content.
Standard item-ranking protocols evaluate candidates independently, which often generates redundant slates that offer low marginal utility.
Given these limitations, it remains underexplored how to systematically integrate evidence-grounded user state modeling with role-aware planning and complementary slate curation.
This gap hinders progress toward recommender systems that operate natively across a user's digital environment.

In this paper, we present \textbf{ClawRec}, the first recommender system designed to operate natively in an agent-accessible user environment.
Unlike traditional recommenders that rely on implicit user state representations, ClawRec establishes an evidence-linked, transparent user state as an explicit bridge between cross-platform behaviors and cross-source recommendations.
To resolve signal heterogeneity, unified event construction standardizes multi-platform behaviors into a comparable format while preserving their underlying semantics, uncertainty, and provenance.
To disentangle temporally mixed histories, user state reasoning dynamically identifies the active task, retains verified persistent preferences, and actively expires or suppresses stale and negatively reinforced signals.
To operationalize this state across sources, role-aware planning translates the active task into concrete support objectives, directing retrieval to sources best suited for specific functional roles.
Finally, to eliminate slate-level redundancy, marginal curation evaluates candidates based on their incremental utility, producing a recommendation slate where items are collectively complementary across distinct support functions.
Backed by end-to-end provenance links from evidence to feedback, ClawRec ensures continuous state evolution, preventing temporary noise from being misidentified as persistent preferences.

To evaluate ClawRec, we introduce \textbf{ClawRec-SimBench}, a benchmark built around temporally evolving user needs.
Each simulated user is associated with a sequence of concrete life events.
Each event captures an immediate real-world task and is realized as a browsing trajectory across heterogeneous information sources.
At each evaluation point, the system observes the preceding behaviors, then recommends content that supports the held-out next behavior.
The benchmark evaluates two aspects of system performance: the quality of the displayed recommendations and the quality of the inferred user state.
ClawRec achieves an NDCG@20 of 0.6134 and a Hit@20 of 0.6944, outperforming the strongest baselines by 0.1126 and 0.0854, respectively.
Its inferred user states also identify the active task more accurately, separate stale history more effectively, and produce fewer unsupported conclusions.
Ablation results further show the contributions of unified event construction, temporal state management, source extension, and marginal curation.

Our main contributions are as follows:

$\bullet$ This paper introduces the Claw-native recommendation setting and presents \textbf{ClawRec}, which turns heterogeneous cross-platform behavior into an evidence-linked user state and uses that state to produce complementary next-step recommendations from functionally appropriate sources.

$\bullet$ To make this setting evaluable, we construct \textbf{ClawRec-SimBench}, a benchmark built from sequences of concrete life events and cross-platform behavior trajectories. Independent human evaluation further validates the quality of its trajectories.

$\bullet$ Extensive experiments on ClawRec-SimBench validate the effectiveness of ClawRec. ClawRec improves both user need inference and recommendation quality over existing methods. These results demonstrate the value of Claw-native recommendation: it moves beyond platform-local ranking toward user-centered recommendation across the broader digital environment.
\section{Approach}

\subsection{Overview of the Approach}

\paragraph{Task Formulation.}
ClawRec considers task-level next-step recommendation from heterogeneous behavioral evidence.
We define the \emph{current task} as the temporally active information-seeking or decision-making objective inferred from a user's recent behavior.
For user \(u\) at decision step \(t\), let \(\mathcal{B}_{t}\) denote the behaviors newly observed since the previous decision, \(S_{t-1}\) denote the previous user state, and \(F_{t-1}\) denote the feedback on the recommendations presented at step \(t-1\).
ClawRec updates the user state \(S_{t}\) from these inputs, and then produces an ordered recommendation slate
\[
\mathcal{R}_t = \left(r_{t,1}, r_{t,2}, \ldots, r_{t,n}\right),
\]
where each \(r_{t,i}\) denotes a recommended item and \(n\) is bounded by a fixed display budget.
Here, an item is a retrieved content unit, such as a web page, video, or product.
The ranking objective is to place recommendations with greater expected usefulness for supporting the current task earlier in the list.
Section~\ref{sec:case_study} presents a case execution trace.

\begin{figure}[t]
    \centering
    \includegraphics[width=0.99\linewidth]{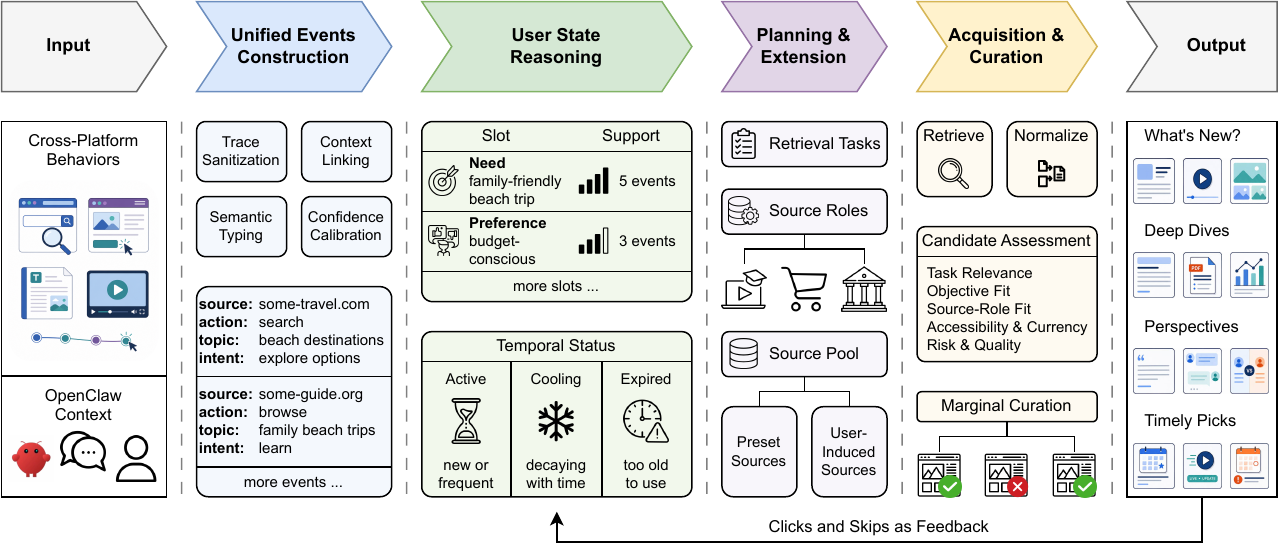}
    \caption{System overview of ClawRec. Cross-platform behaviors and explicit OpenClaw context are converted into unified events, committed to typed and evidence-linked user state records with temporal status, compiled into role-aware retrieval tasks, and curated into recommendation slates, with clicks and skips closing the update loop.}
    \label{fig:overview}
\end{figure}

\paragraph{System Overview.}
To realize the Claw-native recommendation paradigm, we propose ClawRec, a recommender system that operates natively in an agent-accessible user environment.
As illustrated in Figure~\ref{fig:overview}, ClawRec maintains an evidence-linked user state as the interface between heterogeneous behavioral evidence and cross-source recommendation.
To solve preference and state modeling over heterogeneous, cross-platform behaviors, ClawRec first standardizes multi-source interaction traces $\mathcal{B}_t$ (e.g., searches, content visits, and comparisons) into provenance-linked unified events $\mathcal{E}_t$ (Section~\ref{sec:unified_event}).
User state reasoning then consolidates semantically consistent evidence across sources, separating task-specific signals from persistent preferences and source-role knowledge (Section~\ref{sec:user_state_reasoning}).
The resulting state \(S_t\) exposes the active task, supported constraints, and relevant source roles needed for recommendation.
ClawRec uses this state to formulate bounded retrieval objectives over functionally appropriate sources (Section~\ref{sec:plan_extend}), and curates the retrieved candidates according to both their individual utility and their marginal contribution to the final slate (Section~\ref{sec:acquire_curate}).
The produced slate \(\mathcal{R}_t\) therefore combines complementary forms of support rather than redundant items from the same topical or platform context.
Feedback \(F_t\) subsequently revises the evidential support of the corresponding state records and retrieval directions.
Throughout this process, provenance links connect behaviors, state updates, and recommendations, making the system's preference modeling and recommendation decisions traceable.

\subsection{Unified Event Construction}
\label{sec:unified_event}

Raw browser traces are not directly comparable as task evidence.
The same current task may appear through searches, content visits, comparisons, and revisits across sources, while incidental, navigational, or privacy-sensitive pages may provide little usable support.
ClawRec therefore represents each observable behavior as a normalized evidence record before deciding how it should affect the user state.

\paragraph{Event Representation.}
A \emph{unified event} is a normalized, provenance-linked evidence record anchored to one observable behavior.
The behavior segment \(\mathcal{B}_{t}\) is converted into unified events \(\mathcal{E}_{t}\).
Session membership and query-to-content chains link related events and provide local interpretive context, but they do not collapse multiple behaviors into a single event.
Each unified event contains provenance, semantic, and governance information.
Provenance fields record when and where the behavior occurred, its normalized action type, its canonical content target, and an evidence handle for subsequent audit.
Semantic fields record the topic and the behavior-level intent expressed by the action.
Governance fields record confidence and privacy status, which constrain how strongly the event can affect downstream state updates.
The normalized action type abstracts over source-specific behaviors such as search, content consumption, comparison, and revisit.
The behavior-level intent describes the local purpose of the action rather than the complete current task.
The value \texttt{unknown} is retained as an explicit abstention value when the available evidence does not support a more specific intent.

\paragraph{Evidence Preparation.}
ClawRec first sanitizes the observable traces before semantic inference.
It canonicalizes URLs, removes tracking parameters, maps domains to sources, redacts sensitive entities, and filters sensitive or purely navigational pages.
The remaining behaviors are linked to sessions and query-to-content chains.
This local structure allows a query and its subsequent content visits to provide mutually supporting context while preserving a separate evidence handle for each behavior.

\paragraph{Semantic Typing and Calibration.}
On the sanitized and contextualized traces, ClawRec infers a topic and behavior-level intent from observable fields such as query text, title, source, page type, and session context.
Rule-based extraction is used when the source schema provides sufficient information;
otherwise, an LLM is invoked under a constrained schema matching the unified-event interface.
Confidence reflects both the reliability of the observed action and the strength of its contextual support.
Explicit searches, focused comparisons, revisits, meaningful dwell time, and coherent query-to-content chains provide stronger evidence, whereas semantic ambiguity, weak session connection, privacy risk, and page-quality concerns reduce confidence.
An event is emitted only when its required provenance and governance fields are available.
Low-confidence events may provide limited support for the current task, but they do not by themselves justify a reusable preference.

Unlike direct prompting over unstructured history text, unified events separate observable behavior, semantic interpretation, uncertainty, and provenance.
This normalized evidence interface makes heterogeneous behaviors comparable and keeps downstream decisions traceable to specific evidence handles.

\subsection{User State Reasoning}
\label{sec:user_state_reasoning}

Unified events make heterogeneous behaviors comparable, but observation does not imply persistence.
A behavior may provide evidence for the current task, reinforce a reusable constraint, suggest that an earlier task is ending, or indicate that a retrieval direction should be deprioritized.
User state reasoning determines what each event supports, with what confidence, and for how long.
Unlike undifferentiated history accumulation, ClawRec does not make evidence durable merely because it has been observed.

\paragraph{State Representation.}
To prevent evidence about the current task, reusable context, and source usefulness from collapsing into a single undifferentiated memory, ClawRec represents \(S_t\) as evidence-linked state records.
The machine-readable state separates task slots, preference slots, and source-role records.
A task slot represents a concrete information-seeking or decision-support task.
A preference slot stores a reusable preference or constraint only after it receives repeated or explicit support.
A source-role record represents evidence that a source can provide a particular function in a relevant task context.
Negative feedback is attached to the affected records or planning directions as provenance-linked support and constraints rather than stored as a separate slot type.
Each record contains semantic fields, control fields, evidence fields, and provenance fields.
These fields identify what the record represents, its temporal status, positive and negative support, confidence, observation times, and the event or feedback handles that justify it.
ClawRec also maintains a synchronized editable summary so that users can inspect or correct the structured state.

\paragraph{State Update.}

ClawRec revises the state through evidence-referenced patches rather than regenerating it from an unstructured history:
\[
P_t =
\operatorname{Patch}
\left(
S_{t-1},
\mathcal{E}_{t},
F_{t-1}
\right),
\qquad
S_t =
\operatorname{Apply}
\left(
S_{t-1},
\operatorname{Validate}(P_t)
\right).
\]
When explicit user-provided context is available, it is included in the patch request and may directly support or correct the affected records;
it is omitted from the notation for brevity.
New events are linked to existing records according to topic, behavior-level intent, entity continuity, source context, and session continuity.
When no supported match exists, the reasoner creates a new task slot with limited support rather than immediately treating the evidence as a persistent preference.
Linked evidence updates positive and negative support, confidence, timestamps, and provenance.
The patch may create or merge records, revise their support, change temporal status, or apply suppression constraints, with every change justified by event or feedback identifiers.
Before commitment, validation checks the record schema, referenced evidence handles, temporal consistency, and unsupported sensitive inference.
Demographic, health, financial, or family attributes are not committed to the state without explicit user statements or repeated reliable evidence.

\paragraph{State Lifecycle.}
User state records should not retain the same influence as their evidence ages, the task changes, or negative feedback accumulates.
ClawRec therefore distinguishes temporal relevance from feedback-induced suppression:
\[
\mathrm{active}
\rightarrow
\mathrm{cooling}
\rightarrow
\mathrm{expired},
\qquad
\{\mathrm{active},\mathrm{cooling}\}
\rightarrow
\mathrm{suppressed}.
\]
A task slot is active when it has sufficient recent support and remains relevant to the current trigger.
It moves to cooling when reinforcing evidence has been absent or when the observed trajectory contains cues that the task is approaching completion or resolution.
Cooling records remain available as weak context but should not dominate the current recommendation.
A record becomes expired when its support has decayed below the required level or when newer evidence supports a conflicting task interpretation.
Expired records are excluded from the current retrieval plan unless fresh evidence revives them.
Repeated negative feedback on the same task or source role can instead induce suppression.
Suppression blocks or strongly deprioritizes the negatively reinforced direction until new positive evidence supports revision;
it is distinct from expiration caused by time or task change.

The resulting state exposes the active task to be served, reusable context that may constrain it, and stale or negatively reinforced directions that should not dominate retrieval.
Planning must then determine which functional kinds of information would support that task and which sources can provide them.

\subsection{Planning and Source Extension}
\label{sec:plan_extend}

Identifying the current task does not yet determine what should be retrieved.
A useful recommendation slate may need explanation, practical guidance, comparison, experience-based evidence, concrete options, or authoritative verification, and these functions are not necessarily provided by the source of the most recent behavior.
ClawRec therefore plans from the active task and its evidence-backed constraints, rather than directly expanding the latest query or source.

\paragraph{Planning Representation.}
The central distinction in planning is between the types of support that the recommendation slate should cover and the sources used to provide them.
Behavior-level intents contribute to the active task but do not directly determine the slate organization.
A \emph{support type} specifies a general form of task support, while a \emph{support objective} instantiates that type for the current task.
A \emph{support group} is the subset of recommended items in the final slate that jointly fulfills one support objective.
A \emph{source role} describes the function that a source can credibly provide in a particular task context.
ClawRec stores such associations as source-role records containing the source, supported role, task-context anchors, provenance, confidence, retriever capabilities, and temporal status.
The retrieval plan is represented as a set \(\mathcal{T}_t\) of retrieval tasks.
Each \(\tau\in\mathcal{T}_t\) operationalizes a support objective by specifying its support type, source selection, retrieval configuration, and state provenance.

\paragraph{Role-Aware Planning.}
The active task determines which forms of support the slate should cover, while supported user preferences, temporal status, and suppression signals delimit admissible retrieval directions.
Support types and task-specific objectives make these required forms of support explicit.
Source-role records, source permissions, retriever capabilities, and resource budgets then restrict which sources and queries can realize each objective.
The planner records these decisions in \(\mathcal{T}_t\), together with the relevant entities, scenario conditions, language, recency, price, content-type, and risk constraints.
Retrieval expansion is therefore task-directed and bounded rather than an unconstrained continuation of recent browsing.

\paragraph{Behavior-Adaptive Source Extension.}
A fixed allowed-source list provides a controlled retrieval boundary, but it can miss sources that repeatedly prove useful for a particular function in the user's task history.
ClawRec therefore augments the preset list with an evidence-backed, user-induced source set.
A source enters the user-induced set only when repeated evidence supports a stable association between that source and a useful role in relevant task contexts.
Evidence for this association may include repeated successful use, positive feedback on recommendations, recurring transitions from similar tasks to the source, and consistent use of the source for the same role.
Frequent visits alone do not establish a stable source-role association.
For example, repeated use of a community forum for product troubleshooting can support an experience-reference role, while recurring transitions from travel notes to official venue pages can support an official-verification role.
Source-pool membership neither bypasses source permissions nor guarantees selection in the current plan: the active task, required support type, confidence, suppression constraints, and resource budget still determine whether the source is used.
Source-role evidence therefore expands the sources available to planning without treating source use as a persistent user preference.

ClawRec separates the function to be served from the source used to serve it.
This separation allows the system to consider behavior beyond the most recent interactions while keeping retrieval focused on the task, grounded in evidence, and constrained by permissions.

\subsection{Content Acquisition and Marginal Curation}
\label{sec:acquire_curate}

A retrieval plan specifies what information to seek and which sources to query, but it does not guarantee executable retrieval or a useful recommendation slate.
Source-specific queries may be unsupported, returned pages may be inaccessible or duplicated, and individually relevant results may add little value once the slate already contains similar content.
These failures arise at different decision levels: execution feasibility depends on the source interface, individual utility depends on the candidate item and its assigned support objective, and complementarity depends on the partially selected slate.
ClawRec handles these levels through source-aware acquisition, candidate item assessment, and slate-level marginal selection.

\paragraph{Retriever-Aware Acquisition.}

Because sources expose different query languages, operators, and result schemas, a planned retrieval task cannot be executed through a single source-agnostic interface.
ClawRec therefore uses a source adapter to map each \(\tau\in\mathcal{T}_t\) to the target interface.
The adapter selects a source-compatible query candidate, executes it under the task budget, and returns source-specific results.
Each \emph{raw retrieval result} retains the planned instruction, executed query, target source, source role, and consumed budget as retrieval provenance.
Exact URL duplicates, canonical duplicates, and repeated retrieval records are removed before assessment.
When acquisition fails, the failure status is recorded separately from preference feedback.
Recovery actions, such as relaxing optional constraints or trying the next query candidate, consume the remaining retrieval budget.

\paragraph{Candidate Item Representation.}
To assess heterogeneous raw retrieval results under a common curation policy, ClawRec normalizes each result into a \emph{candidate item}, forming \(\mathcal{C}_t\).
A candidate item is an internal representation used before final selection; it becomes a recommended item only if it is selected into \(\mathcal{R}_t\).
Each candidate item exposes the information needed for utility assessment, provenance tracking, and risk control through four groups of fields.
Content fields include the title, URL, source, content type, and a short summary.
Provenance fields link the candidate item to the executed query, retrieval task, and state records that motivated it.
Planning fields record the intended support type, support objective, and source role.
Risk fields record privacy, accessibility, availability, source-quality, and execution concerns.

\paragraph{Item-Level Assessment.}
ClawRec first removes candidate items that violate retrieval constraints, privacy rules, accumulated suppression signals, accessibility requirements, or minimum evidence requirements.
Each remaining candidate item is assessed for utility with respect to the current state \(S_t\), its retrieval task \(\tau\), and the relevant feedback constraints.
The assessment evaluates the item's relevance to the active task, fit with the assigned support objective, and compatibility with the intended source role.
It also considers whether the item is sufficiently current and accessible and whether its use introduces unsupported or excessive risk.
Privacy violations, task conflicts, suppression conflicts, and severe execution risks are hard removal conditions.
Weak task match, broad topicality, uncertainty, and moderate source-quality concerns act as soft penalties.
When otherwise comparable, candidate items with stronger provenance, better support-type and source-role fit, lower risk, and higher planned priority are preferred.

\paragraph{Slate-Level Marginal Selection.}
After candidate-item assessment, ClawRec performs diversity-aware marginal selection~\cite{carbonell1998use}.
The slate-level marginal value of an eligible candidate item is the additional task support it contributes relative to the items already selected for the slate.
Starting from an empty slate and operating under the shared display budget, ClawRec considers the planned support objectives and repeatedly selects the candidate item with the highest marginal value.
A candidate item has high marginal value when it covers an under-served support type, contributes a distinct source role, adds actionable information, or supplies trusted confirmation that complements the selected items.
Its marginal value is low when it semantically repeats existing content, duplicates an already covered function, only broadly matches the topic, reinforces an expired task, or adds risk without additional decision support.
If the available candidates cannot fulfill a planned support objective, ClawRec leaves the corresponding support group empty rather than filling it with weakly relevant content.
The selected candidate items become the displayed recommended items and form the recommendation slate \(\mathcal{R}_t\), whose organization reflects item utility within each support group and complementarity across the slate.

\paragraph{Feedback Provenance.}
After \(\mathcal{R}_t\) is displayed, feedback on each recommended item is linked to the task slot, source-role record, retrieval task, and candidate item from which it was selected.
A click provides positive support because the item was useful enough to inspect.
A skip provides only weak negative support because it may reflect poor relevance, duplicate coverage, timing, or limited attention.
A single skip is therefore not treated as a stable dislike.
Repeated negative feedback on the same task, retrieval direction, or source role may accumulate into a suppression constraint for later planning and source selection.
Such feedback constrains a retrieval direction rather than establishing that the corresponding user preference is absent.

This separation prevents three distinct failure modes from being conflated: source-side execution failure, weak candidate-item utility, and poor slate-level complementarity.
Provenance links retained across events, state records, retrieval tasks, candidate assessment, marginal selection, and feedback jointly form the audit trail.
Appendix~\ref{app:controls} describes the corresponding privacy and audit controls, which provide practical transparency and user control rather than formal privacy guarantees.

\section{ClawRec-SimBench: Benchmarking Recommender Agents for Cross-Platform Next-Step Recommendation}
\label{sec:simbench}

Claw-native recommendation introduces an evaluation problem that is not adequately captured by either conventional recommender datasets or standard agent benchmarks.
Recommender datasets typically assume a platform-specific item space and evaluate whether a system ranks or predicts an observed item.
This formulation does not test whether a system can recover a temporally active task from heterogeneous cross-platform behaviors, distinguish it from reusable or stale historical context, and identify content from functionally appropriate sources.
In contrast, Web-agent benchmarks commonly provide the user goal explicitly and evaluate whether an agent can navigate interfaces, use tools, or complete the specified task~\cite{zhou2023webarena,deng2023mind2web,xie2024osworld}.
In the setting studied here, the goal itself is latent: the system must infer the user's current task from partial behavioral evidence before deciding what content would provide useful next-step support.

To represent the latent context that gives rise to this behavior, ClawRec-SimBench uses a \emph{life event}: a concrete real-world situation that creates an immediate information-seeking or decision-support task.
A simulated user is assigned a sequence of life events, and each event is realized as a platform-conditioned, GUI-grounded behavior trajectory across heterogeneous sources.
This design separates the simulator-side task context from the evidence available to the evaluated system.

\begin{figure}[t]
    \centering
    \includegraphics[width=0.95\linewidth]{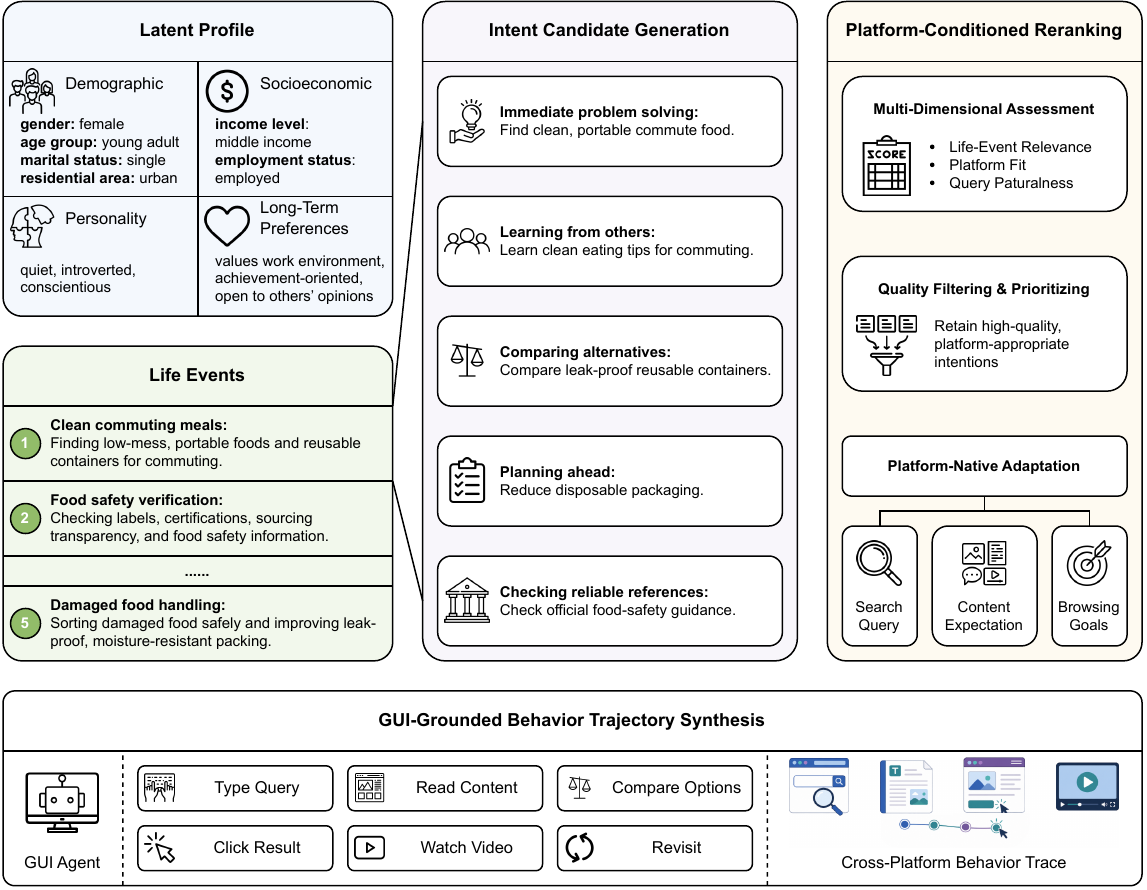}
    \caption{Overview of ClawRec-SimBench construction. Latent profiles and life events define simulator user contexts, which are transformed into diverse browsing intents, filtered and rewritten for specific platforms, and executed by a GUI agent to generate cross-platform behavior traces.}
    \label{fig:simbench_pipeline}
\end{figure}

Evaluated systems observe only the authorized prefix of the current event trajectory and the complete behavior histories of previous life events.
From this evidence, they must infer the current user state and produce ranked recommendations that support a held-out next behavior.
Because multiple pages or sources may provide useful support for the same current task, the benchmark evaluates task-level semantic usefulness rather than exact URL or item-identity prediction.
Figure~\ref{fig:simbench_pipeline} summarizes how latent user context is transformed into platform-aware browsing behavior and grounded cross-platform trajectories for this evaluation.

\subsection{Synthetic User Generation}

Each simulated user \(u\) is associated with a latent profile \(P^{(u)}\).
The profile contains demographic, socioeconomic, personality, and long-term preference information.
The simulator uses this information solely to generate behavior and evaluate state consistency.
ClawRec never observes \(P^{(u)}\) at inference time.

User \(u\) is paired with a sequence of synthetic life events:
\[
L^{(u)} = \left(\ell^{(u)}_1, \ell^{(u)}_2, \ldots, \ell^{(u)}_{M}\right),
\]
where \(M\) is the number of life events, and, for \(m\in\{1,\ldots,M\}\), each \(\ell^{(u)}_m\) is a concrete situation that gives rise to an immediate information-seeking or decision-support task.
In the current benchmark design, each user is assigned multiple consecutive life events.
These events may share persistent user constraints or a broader context while corresponding to different immediate tasks.
For example, a user preparing for a multi-day trip may first look for durable food containers, later plan portable low-sodium meals under a dietary constraint, and, after the trip, seek guidance on handling luggage contamination.

This structure tests whether a system can transfer supported reusable context across events while grounding each current task separately and preventing completed tasks from dominating later recommendations.
It also separates simulator-side latent variables from the evidence available to ClawRec: the latent profile \(P^{(u)}\) and life-event sequence \(L^{(u)}\) are used to synthesize behavior, whereas ClawRec observes only authorized behavior traces during inference.

\subsection{Behavior Trajectory Synthesis}

For each life event \(\ell^{(u)}_m\), the simulator synthesizes platform-aware browsing intentions through a two-stage procedure and then executes them through its GUI capability.

\paragraph{Intent Candidate Generation.}
Given the latent profile \(P^{(u)}\), the recent event history, and the current life event \(\ell^{(u)}_m\), the simulator first produces a diverse set of browsing-intention candidates.
These candidates represent different information requirements that may arise in the current event.
Each candidate is annotated with a driver type that explains the underlying motivation of the browsing intention.
The driver types cover multiple information-seeking modes, including solving an immediate problem, learning from others, comparing alternatives, planning ahead, and checking reliable references.
The candidates are grounded in the current life event, recurring historical patterns, or cross-event inference.
At this stage, generation remains platform-agnostic: the candidates contain only weak platform hints and neutral query seed terms, rather than platform-specific search queries or concrete browsing goals.

\paragraph{Platform-Conditioned Reranking.}
For each target platform, the simulator scores every candidate along three dimensions: life-event relevance, platform fit, and query naturalness.
Each dimension is scored on a 1--5 scale.
Candidates with a total score below 9, or below 3 on either platform fit or query naturalness, are filtered out.
Qualified candidates are rewritten into platform-native search queries, content-type expectations, and concrete browsing goals.
All platform-specific candidates are then pooled, and the top 5 by score are retained for each life event.
For example, the life event ``prepare clean, portable meals for commuting'' may be realized as a lifestyle-sharing query for low-mess meal ideas, an e-commerce query for leak-proof reusable containers, and a query to an official information source for food-safety guidance.

\paragraph{GUI-Grounded Behavior Trajectory Synthesis.}
A GUI agent then executes the plan by interacting with actual platform surfaces: it types queries into search boxes, clicks on results, reads content, compares options, watches content, and revisits sources.
The agent does not emit a structured log directly; instead, its behavior is captured as a cross-platform trajectory
\[
B^{(u)}_m
=
\left(
 b^{(u)}_{m,1},
 b^{(u)}_{m,2},
 \ldots,
 b^{(u)}_{m,K}
\right),
\]
where \(K\) is the number of behavior steps generated for life event \(m\), and, for \(k\in\{1,\ldots,K\}\), each \(b^{(u)}_{m,k}\) records the observable platform, action type, and query or content target at step \(k\).
The simulator separately retains the intent that motivated each action as a hidden annotation for trajectory construction and evaluation.

Platform-aware intent realization is central to the benchmark.
The selected platforms cover community discussion, lifestyle sharing, e-commerce, video, search, and official information channels.
This makes the generated behavior traces resemble how a user moves across the Web while addressing the task induced by a concrete life event.

\subsection{Leave-One-Out Evaluation Protocol}

After synthesizing a complete behavior trajectory for each life event, we construct an evaluation instance immediately before its final behavior.
The system receives the complete behavior histories of earlier life events but only a prefix of the current event trajectory.
This setting tests whether it can reuse supported cross-event context while grounding its recommendations in the active task, rather than allowing completed or stale historical tasks to dominate the current decision.
From this partial evidence, the system must infer the current user state and rank content that would provide useful next-step support.

For life event \(m\) of user \(u\), define
\[
\begin{aligned}
B^{(u)}_{<m}
&=
\left(B^{(u)}_1,\ldots,B^{(u)}_{m-1}\right)
&& \text{(previous life-event histories)},\\
B^{(u)}_{m,<K}
&=
\left(b^{(u)}_{m,1},\ldots,b^{(u)}_{m,K-1}\right)
&& \text{(observed prefix of the current event)},\\
h^{(u)}_m
&=
b^{(u)}_{m,K}
&& \text{(held-out next behavior)}.
\end{aligned}
\]
At inference time, an evaluated system receives \(B^{(u)}_{<m}\) and \(B^{(u)}_{m,<K}\), while \(h^{(u)}_m\) remains hidden.
For the first life event, \(B^{(u)}_{<1}=\varnothing\).

The information boundary follows the evidence available before the held-out behavior occurs.
Simulator-only fields, including latent profiles, life-event descriptions, driver types, candidate scores, intent labels, source-role labels, and the held-out behavior, are removed before system inference.
Systems receive only authorized observable fields, including timestamps, actions, queries, page titles or content, and source identifiers.
In particular, source-role assignments must be inferred from behavioral evidence rather than supplied as annotations.

The held-out behavior serves as a natural semantic reference for the next step, not as a unique item that the system must reproduce.
The protocol therefore does not require an exact match to the same URL, page, or product.
A recommendation may still be useful when it supports the same current task, concrete topic, behavior-level intent, or source role through different content.
The corresponding task-level relevance rubric and ranking metrics are defined in the Metrics subsection.

\subsection{Benchmark Statistics and Quality Validation}
\label{sec:benchmark_statistics_validation}

Before using ClawRec-SimBench for system comparison, we summarize its scale, trajectory distribution, and validation quality.
The analysis verifies that the benchmark provides heterogeneous behavior evidence, diverse source-role coverage, and plausible held-out targets for task-level next-step recommendation.

\subsubsection{Benchmark Distribution Analysis}

\begin{table}[t]
\centering
\caption{The benchmark statistics characterize the composition of ClawRec-SimBench, while the validation scores assess its overall quality.}
\label{tab:benchmark_stats_quality}

\begin{subtable}{0.43\linewidth}
\centering
\caption{Benchmark statistics.}
\label{tab:benchmark_stats}
\begin{tabular}{lc}
\toprule
\textbf{Statistic} & \textbf{Value} \\
\midrule
Users & 109 \\
Events & 545 \\
Avg. trajectory length & 28.82 \\
Sources & 166 \\
\bottomrule
\end{tabular}
\end{subtable}
\hfill
\begin{subtable}{0.53\linewidth}
\centering
\caption{Quality validation.}
\label{tab:quality_validation}
\begin{tabular}{lc}
\toprule
\textbf{Metric} & \textbf{Score} \\
\midrule
Trajectory Coherence & 4.52 \\
Query Naturalness & 4.04 \\
Source-Role Plausibility & 4.88 \\
Target Plausibility & 4.53 \\
\bottomrule
\end{tabular}
\end{subtable}

\end{table}

Table~\ref{tab:benchmark_stats} reports the scale of the experimental split.
It contains 109 simulated users, 545 life events, an average trajectory length of 28.82 behavior steps, and 166 distinct sources.
These statistics define the evaluation scope used in the following experiments.

\begin{figure}[t]
    \centering
    \begin{subfigure}{0.48\linewidth}
        \centering
        \includegraphics[width=\linewidth]{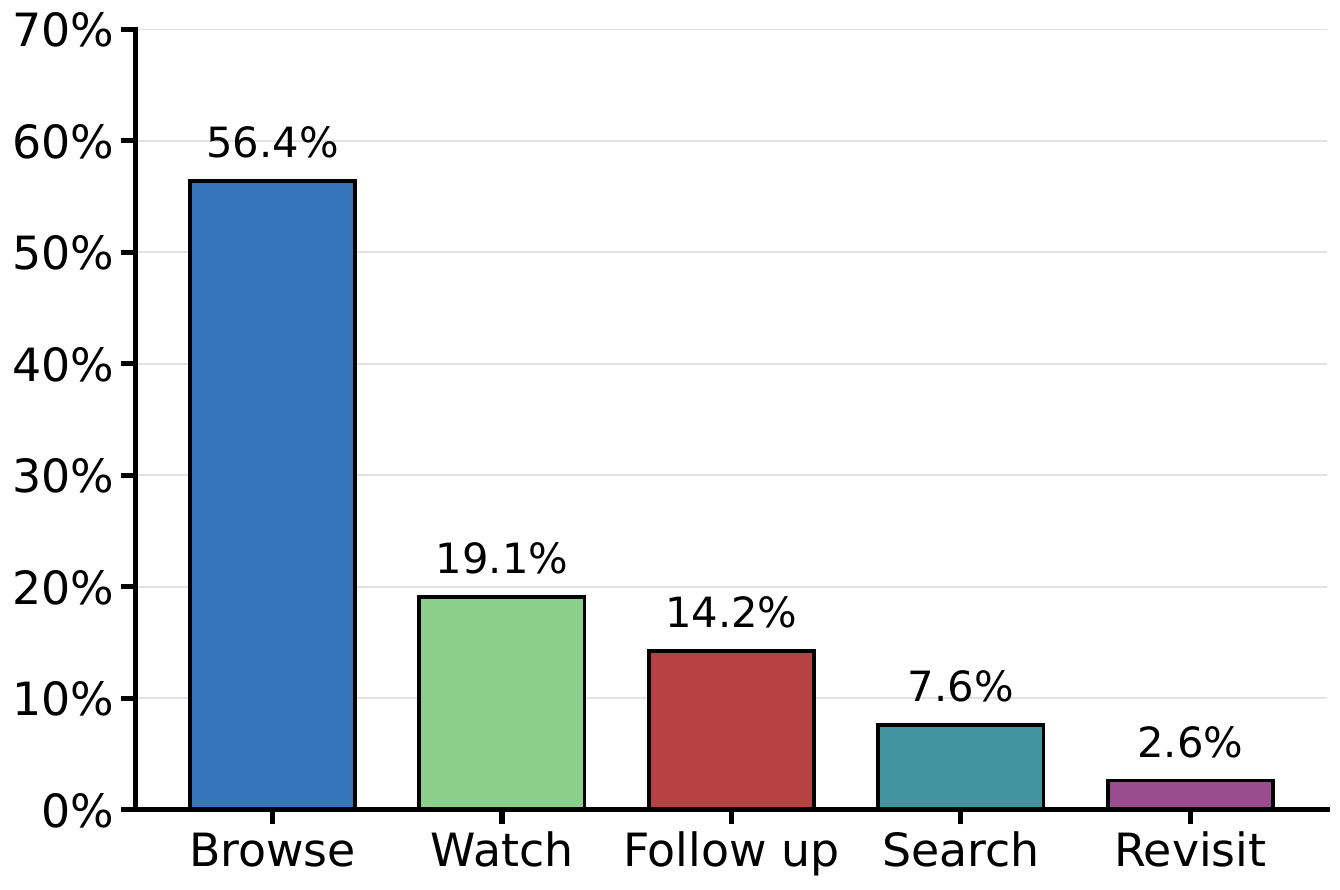}
        \caption{Behavior action distribution.}
        \label{fig:behavior_action_distribution}
    \end{subfigure}
    \hfill
    \begin{subfigure}{0.48\linewidth}
        \centering
        \includegraphics[width=\linewidth]{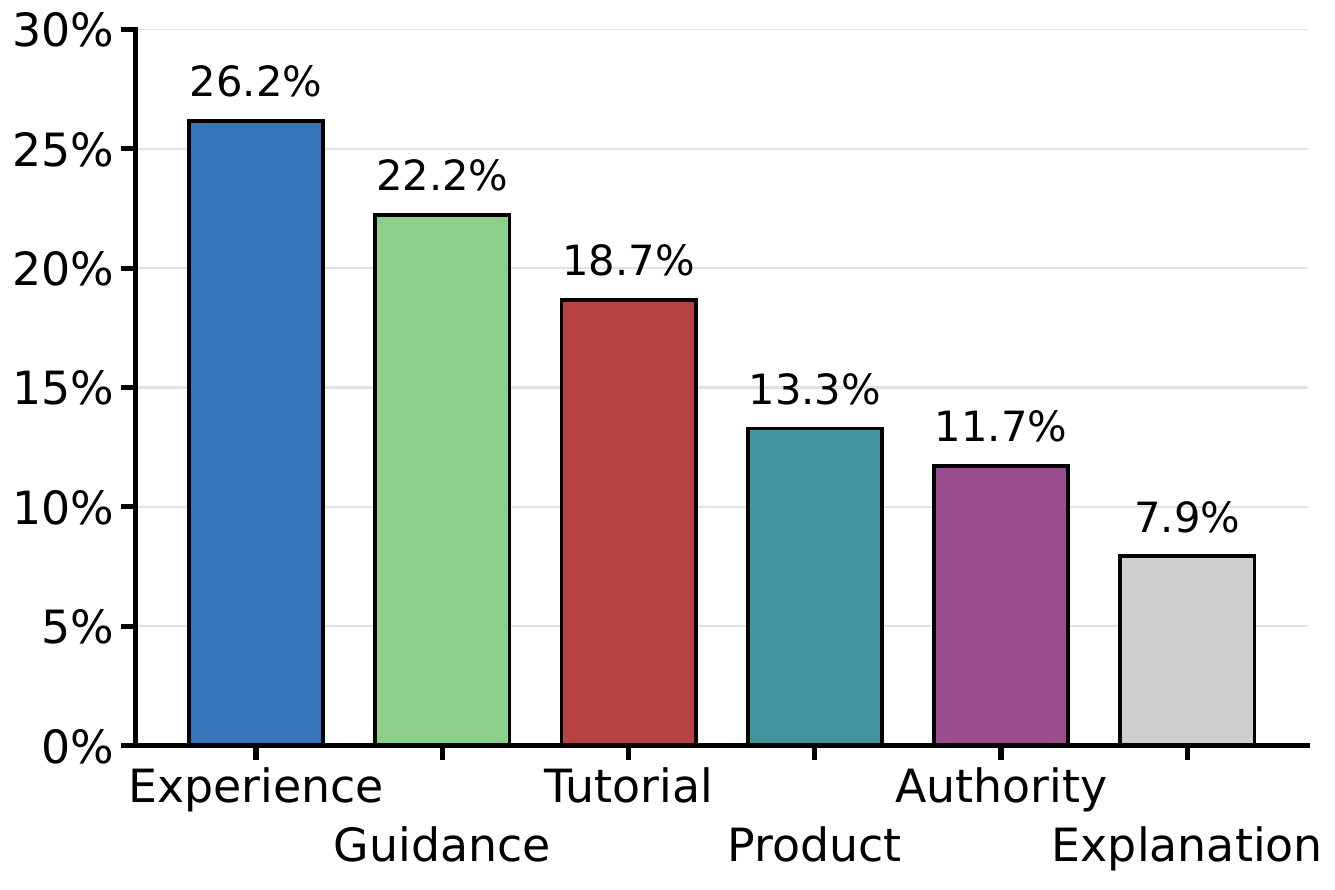}
        \caption{Source-role distribution.}
        \label{fig:source_role_distribution}
    \end{subfigure}
    \caption{Benchmark distribution of simulated user behaviors and source roles. Browse dominates the behavior-action distribution, whereas source roles span multiple functional categories.}
    \label{fig:benchmark_distribution}
\end{figure}

Figure~\ref{fig:benchmark_distribution} provides a distributional view of ClawRec-SimBench along two dimensions: behavior actions and source roles.
The action distribution is dominated by browsing (56.4\%), followed by watching and follow-up actions, while search and revisit behaviors occur less frequently.
In contrast, source roles are more evenly spread across experience, guidance, tutorial, product, authority, and explanation categories.
This indicates that the benchmark contains diverse behavioral traces, together with broad functional coverage of recommendation sources.
This coverage enables evaluation of systems that infer the current task from heterogeneous behavior evidence and recommend content with an appropriate functional role.

\paragraph{Data Split.}
\label{sec:data_split}
ClawRec-SimBench uses a user-level split, with all life events of the same user assigned to one split.
The development set contains 20 users and 100 life events, while the test set contains 89 users and 445 life events.
We use the development set to finalize system configurations, evaluation rubrics, and judge prompts, and report results on the test set.

\subsubsection{Quality Validation}

For quality validation, we randomly sample 50 life events from ClawRec-SimBench.
Two human annotators independently evaluate each sampled event using four metrics: Trajectory Coherence, Query Naturalness, Source-Role Plausibility, and Target Plausibility.
Each metric is scored on a 1--5 scale.
The scores reported in Table~\ref{tab:quality_validation} are averaged over the two annotators and the 50 sampled events.

\paragraph{Trajectory Coherence.}
This metric measures whether the behavior sequence develops around the same underlying life event.
A coherent trajectory presents clear task continuity from initial exploration to later comparison, inspection, watching, or verification.
The score of 4.52 indicates that most generated trajectories preserve a consistent information-seeking process.

\paragraph{Query Naturalness.}
This metric measures whether platform-specific queries resemble realistic user searches.
Natural queries follow the style and granularity of the target platform, rather than appearing as over-specified simulator instructions.
The score of 4.04 indicates that the generated queries are generally natural while leaving room for query-level noise.

\paragraph{Source-Role Plausibility.}
This metric measures whether each source is used for a plausible functional role in the current task.
The score of 4.88 shows that source-role assignments are highly consistent with the intended task functions.

\paragraph{Target Plausibility.}
This metric measures whether the hidden final behavior forms a reasonable next-step target given the observed prefix.
A plausible target is semantically aligned with the current life event and represents a natural continuation of the user's information-seeking process.
The score of 4.53 supports the use of the hidden final behavior as the leave-one-out target for task-level recommendation evaluation.

Overall, the statistics, distribution analysis, and quality validation show that ClawRec-SimBench provides heterogeneous behavior prefixes, diverse source-role targets, and plausible leave-one-out instances.
These properties support the downstream evaluation of contextual intent grounding, user state evolution, and source-role-aware next-step recommendation.

\section{Experiments}
\label{sec:experiments}

We use the ClawRec-SimBench protocol defined in Section~\ref{sec:simbench} and focus here on system comparison and mechanism analysis.
The experiments answer two questions: whether ClawRec improves task-level recommendation quality under matched resource limits, and which system mechanisms account for the improvement.

\subsection{Experimental Setup}

We conduct experiments on ClawRec-SimBench, the benchmark introduced in Section~\ref{sec:simbench}.
Unless otherwise specified, all comparisons follow the matched evaluation setting specified in Implementation Details.

\subsubsection{Baselines}

Recent work has used LLMs for recommendation and ranking, extended recommender systems with agentic planning and memory, and incorporated user histories into personalized Web agents for context-aware Web interaction \cite{wu2024survey,peng2025survey,cai2025large}.
From these lines of work, we select methods that can be adapted to consume the observed behavior history, retrieve Web content at inference time, and return ranked recommended items through a common evaluation interface.
The resulting baselines cover a capability ladder from source-level popularity and query- or history-based retrieval to direct LLM reasoning, personalized Web search, and agentic memory and search.

For a controlled comparison, all LLM-based and agent-based baselines use the same LLM backbone and decoding setting as ClawRec, so that performance differences mainly reflect differences in system design rather than backbone model capacity.
All displayed recommended items are grounded in retrieved or browsed Web content.
For methods without native browser-operation interfaces, we execute their generated queries through a lightweight browser-based retrieval utility implemented with Chrome DevTools MCP \cite{chrome-devtools-mcp} and normalize the returned pages into recommended items for evaluation.

\textbf{(1) Non-LLM methods:}

$\bullet$ \textbf{Popular / Trending} retrieves popular, trending, or highly ranked content from the allowed source pool and ranks the resulting items by source-level popularity signals.

$\bullet$ \textbf{Last-query Retrieval} uses the most recent query in the observed prefix to retrieve source-compatible pages and ranks the returned items by retrieval order.

$\bullet$ \textbf{History-QE + BM25 Rerank} applies pseudo-relevance-feedback query expansion \cite{buckley1994automatic} using the current-event prefix and historical behavior trace, then reranks the retrieved items with BM25 \cite{robertson1994okapi}.

\textbf{(2) Direct LLM reasoning methods:}

$\bullet$ \textbf{LLM Planner + LLMRank} directly prompts an LLM with the raw observed behavior history as text, including previous events and the current-event prefix, to generate retrieval directions.
It collects candidate pages according to these directions and applies LLMRank \cite{hou2024large} to the resulting items.

\textbf{(3) Agentic methods:}
For agentic baselines, we adapt recent systems to the ClawRec-SimBench setting and convert their retrieved evidence or planned next actions into ranked recommended items under the shared display budget.

$\bullet$ \textbf{Persona2Web-style Web Agent} is a task adaptation of the personalized Web-agent setting in Persona2Web \cite{kim2026persona2web}.
The agent receives the observed behavior history and the current-event prefix, issues multi-source searches under the shared retrieval budget, and ranks recommended items derived from its retrieved evidence and planned next actions.
    
$\bullet$ \textbf{MARS-style Memory Agent} adapts the hierarchical belief-state memory design of MARS \cite{shen2026agentic} to ClawRec-SimBench.
Previous life-event behaviors are summarized into event-level and preference-level memories, which are provided to the LLM planner together with the current-event prefix.
The agent then retrieves and ranks candidate items under the same retrieval and display budgets.
    
$\bullet$ \textbf{OpenClaw w/o ClawRec} serves as a generic OpenClaw control \cite{openclaw2026}.
It uses the OpenClaw agent context and conversation memory to plan retrieval and rank items under the shared budgets, with ClawRec-specific recommendation modules disabled.

Together, these baselines compare ClawRec's recommendation-specific state, planning, and curation against generic retrieval, direct LLM reasoning, and agentic memory and search.

\subsubsection{Metrics}

We evaluate ClawRec from two aspects: user state quality and recommendation quality.
The former examines the user state defined in Section~\ref{sec:user_state_reasoning}, while the latter examines whether the displayed recommended items support the user's held-out next-step behavior.

\paragraph{User State Quality.}

We use State Consistency Score (SCS) to evaluate whether the inferred user state is supported by authorized behavioral evidence and aligned with the current life event.
SCS is computed as the arithmetic mean of four state-quality dimensions: Current Need, Preference, Temporal Alignment, and Calibration.
Each dimension is scored on a 1--5 scale, with higher scores indicating stronger evidence support and better alignment with the current life event.

$\bullet$ \textbf{Current Need} measures whether the inferred state identifies the main need behind the current behavior prefix. A high score requires the state to consolidate heterogeneous source evidence into a concrete current task, rather than flattening the prefix into a broad topical interest, a generic shopping intent, or a stale historical need.

$\bullet$ \textbf{Preference} measures whether reusable preferences or constraints are recorded only when they are supported by sufficient evidence. A short behavior chain should not be promoted to a stable preference unless it is supported by repeated behavior, cross-event evidence, or explicit user-provided context.

$\bullet$ \textbf{Temporal Alignment} measures whether the inferred state keeps current needs separated from completed or stale historical needs.

$\bullet$ \textbf{Calibration} measures whether the inferred state expresses uncertainty appropriately and avoids unsupported over-inference. In this dimension, sensitive demographic, health, financial, or family attributes are not treated as targets to be recovered from behavior traces. Inferring such attributes without explicit user statements or repeated reliable evidence is counted as a calibration error.

The final SCS is the average of the four dimensions:
\[
\mathrm{SCS}=
\frac{1}{4}
(s_{\mathrm{need}}+s_{\mathrm{pref}}+s_{\mathrm{temp}}+s_{\mathrm{calib}}).
\]

\paragraph{Recommendation Quality.}

Recommendation quality evaluates whether the displayed recommended items support the user's held-out next-step behavior and whether more useful items are ranked earlier.
Under the protocol defined in Section~\ref{sec:simbench}, the final behavior of each life event is hidden from the system.
We denote the held-out behavior for life event \(m\) of user \(u\) by \(h^{(u)}_m\).
The system produces a ranked recommendation slate from the observed current-event prefix and previous life-event histories.
Unlike standard next-item recommendation based on exact item-ID matching, the objective is not to reproduce the identical URL, page, or product ID.
A recommendation can still be useful if it supports the same current task, behavior-level intent, topic, or source role as the held-out behavior.
We therefore assign task-level relevance scores rather than binary exact-match labels.
For a fixed evaluation instance, let \(r_i\) denote the recommended item displayed at rank \(i\).
Each item receives a relevance score \(\operatorname{rel}(r_i,h^{(u)}_m)\) with respect to the held-out behavior.
The relevance score uses a 1--5 scale: 1 indicates that the item is unrelated or unsupported, 3 indicates partial support for the same current task, and 5 indicates that the item can closely substitute for or directly support the held-out next step.
Based on these task-level relevance scores, we report NDCG@\(K\)~\cite{jarvelin2002cumulated} and Hit@\(K\).
NDCG@\(K\) measures whether highly relevant items are ranked earlier:
\begin{equation*}
\begin{aligned}
\mathrm{DCG}@K
&=
\sum_{i=1}^{K}
\frac{
2^{\operatorname{rel}(r_i,h^{(u)}_m)}-1
}{
\log_2(i+1)
},
\\
\mathrm{NDCG}@K
&=
\frac{\mathrm{DCG}@K}{\mathrm{IDCG}@K}.
\end{aligned}
\end{equation*}
For each test instance, \(\mathrm{IDCG}@K\) is computed from a shared evaluation pool containing the top-\(K\) recommended items produced by all evaluated methods.
The items in this pool are scored using the same task-level relevance rubric, and the resulting ideal ranking is used to compute \(\mathrm{IDCG}@K\).
Each method's \(\mathrm{DCG}@K\) is still computed according to its own displayed order, so NDCG@\(K\) measures ranked task support within the shared evaluation pool.
Hit@\(K\) measures whether the top-\(K\) displayed items contain at least one recommendation that is sufficiently useful for the held-out behavior:
\begin{equation*}
\mathrm{Hit}@K
=
1
\left[
\max_{1 \le i \le K}
\operatorname{rel}(r_i,h^{(u)}_m)
\ge
\tau
\right].
\end{equation*}
Here, \(\tau\) is the relevance threshold for a successful hit.
We set \(\tau=3\), corresponding to partial support for the same current task under the 1--5 task-level relevance rubric.
We report all top-\(K\) recommendation metrics at \(K=20\).

\paragraph{Diagnostic Measures.}

We report diagnostic measures that characterize mechanism-level failures across systems.
These measures support interpretation of current task grounding, self-evolution across events, and source-role-aware retrieval.

$\bullet$ \textbf{Source-Role Match} measures whether the recommended content comes from a source with an appropriate functional role for the current task.

$\bullet$ \textbf{Topic Match} measures whether the recommended content matches the concrete topic of the current life event.
For example, low-sodium plant-based bento preparation requires content specific to that task, not generic healthy-eating material.

$\bullet$ \textbf{Intent Match} measures whether the recommended content supports the same functional intent as the held-out behavior.

\paragraph{Judging Protocol.}

The following judging protocol is used for metrics that require semantic assessment, including the four state-quality dimensions, the task-level relevance scores \(\operatorname{rel}(r_i,h^{(u)}_m)\) used by NDCG@20 and Hit@20, and the diagnostic measures Source-Role Match, Topic Match, and Intent Match.
For each evaluated instance, judges are given only the observed behavior prefix, the relevant system output, the held-out behavior when required by the metric, and the scoring rubric.
They are blind to the method name, the hidden user profile, the event-generation prompt, and system-internal artifacts.
For recommendation quality metrics, the top-\(K\) items from all evaluated methods are placed into a shared evaluation pool for each test instance and scored using the same task-level relevance rubric.
Judges are blind to the method that produced each item.
The resulting relevance scores are mapped back to each method's displayed order for computing DCG@\(K\), NDCG@\(K\), and Hit@\(K\).
On the development split, which contains 20 users, we obtain both human and LLM judge~\cite{zheng2023judging} scores for the evaluated instances.
We use these paired scores to calibrate the LLM judge at the rubric-instruction level~\cite{liu2023geval}, refining the judge prompt and few-shot examples to clarify the boundaries between exact support, semantic substitution, partial task usefulness, broad topical relatedness, and unsupported inference.
The calibrated judge prompt and few-shot examples are then frozen and applied unchanged to the test users.
All reported semantic-assessment metrics on the test set are computed from the LLM judge's scores.

\subsubsection{Implementation Details}

For recommendation generation, ClawRec and all LLM-based baselines use \texttt{DeepSeek-V4-Flash}~\cite{deepseek-ai2026deepseekv4} as the backbone model.
All methods are given the same observed evidence.
The retrieval budget is limited to 10 search tasks, and the display budget is limited to 20 recommended items.
For all recommendation-generation LLM calls, we use the same LLM configuration: the temperature is fixed to 1.0, and the maximum context length is set to 128K tokens.
For semantic evaluation, we use \texttt{GPT-5.5}~\cite{openaiGPT55SystemCard2026} as the LLM judge.
The same temperature and context-length settings are used for these evaluation-stage LLM calls.
The evaluation prompts for state quality, recommendation relevance, and diagnostic metrics are provided in Appendix~\ref{app:evaluation_prompts}.

\subsection{Overall Performance}

Under this matched setup, Table~\ref{tab:main} compares how different systems use the same observable cross-source evidence to infer the active task, manage historical context, and produce ranked next-step recommendations.

\begin{table}[t]
\centering
\small
\caption{
User state modeling and next-step recommendation performance across non-LLM, direct LLM, and agentic methods.
The best and second-best results are highlighted in \textbf{bold} and \underline{underlined}, respectively; ``--'' indicates that SCS is not applicable because the method does not produce an inspectable user state.
}
\renewcommand{\arraystretch}{1.12}
\begin{tabular}{@{}lcccccc@{}}
\toprule
\multirow{2}{*}{\textbf{Method}}
& \multicolumn{4}{c}{\textbf{State Quality}}
& \multicolumn{2}{c}{\textbf{Recommendation Quality}} \\
\cmidrule(lr){2-5}
\cmidrule(lr){6-7}
& \textbf{Need}
& \textbf{Pref.}
& \textbf{Temp.}
& \textbf{Calib.}
& \textbf{NDCG@20}
& \textbf{Hit@20} \\
\midrule

\rowcolor{gray!12}
\multicolumn{7}{c}{\textbf{Non-LLM Methods}} \\
\addlinespace[1pt]
Popular / Trending
& -- & -- & -- & -- & 0.0112 & 0.0135 \\
Last-query Retrieval
& -- & -- & -- & -- & 0.2002 & 0.2315 \\
History-QE + BM25 Rerank
& -- & -- & -- & -- & 0.2482 & 0.3528 \\

\addlinespace[2pt]
\rowcolor{gray!12}
\multicolumn{7}{c}{\textbf{Direct LLM Reasoning Methods}} \\
\addlinespace[1pt]
LLM Planner + LLMRank
& -- & -- & -- & -- & 0.3438 & 0.5393 \\

\addlinespace[2pt]
\rowcolor{gray!12}
\multicolumn{7}{c}{\textbf{Agentic Methods}} \\
\addlinespace[1pt]
Persona2Web-style Web Agent
& 3.52 & 3.14 & 2.88 & 2.26
& 0.4175 & \underline{0.6090} \\
MARS-style Memory Agent
& \underline{4.03} & \textbf{4.05}
& \underline{4.20} & \underline{3.40}
& 0.4404 & 0.5618 \\
OpenClaw w/o ClawRec
& 3.71 & 3.91 & 3.66 & 3.06
& \underline{0.5008} & 0.5730 \\

\midrule
\textbf{ClawRec}
& \textbf{4.57}
& \underline{3.98}
& \textbf{4.29}
& \textbf{3.65}
& \textbf{0.6134}
& \textbf{0.6944} \\
\bottomrule
\end{tabular}
\label{tab:main}
\end{table}

ClawRec achieves the best recommendation quality on both ranking metrics.
It obtains an NDCG@20 of 0.6134, improving over the strongest baseline on this metric, OpenClaw w/o ClawRec, by 0.1126.
It also obtains a Hit@20 of 0.6944, improving over the strongest baseline on this metric, Persona2Web-style Web Agent, by 0.0854.
The NDCG@20 gain indicates that ClawRec places task-supporting items earlier in the displayed recommendation slate, while the Hit@20 gain indicates that it more reliably includes useful next-step content within the fixed display budget.

ClawRec also achieves the highest scores on Current Need, Temporal Alignment, and Calibration.
Its Current Need score of 4.57 reflects stronger recovery of the active task from the observed behavior prefix.
Its Temporal Alignment score of 4.29 reflects better separation of the current need from completed or stale historical needs, while its Calibration score of 3.65 reflects more conservative treatment of weak or unsupported evidence.
ClawRec also remains competitive on Preference, obtaining 3.98 compared with the best score of 4.05.
The state-quality results therefore locate its main advantage in grounding the temporally active task without losing substantial ability to retain supported reusable context.

The baseline pattern further clarifies the difficulties introduced by fragmented cross-platform evidence.
Popular / Trending provides little support for individualized next-step needs because it does not condition its ranking on the observed behavioral trajectory.
Last-query Retrieval and History-QE + BM25 Rerank improve substantially, indicating that recent queries and historical traces contain useful task signals.
Their remaining gap from the LLM-based and agentic methods, however, suggests that local query continuation and lexical aggregation do not fully recover the latent task expressed across heterogeneous behaviors.
LLM Planner + LLMRank improves further by reasoning semantically over the raw behavior history, while the agentic baselines benefit from memory abstraction and multi-step Web search.
Nevertheless, none of these methods achieves the same combination of Current Need, Temporal Alignment, Calibration, and recommendation quality as ClawRec.
This pattern is consistent with the need for recommendation-specific organization of heterogeneous evidence, rather than relying only on additional history, direct prompting, or generic Web-agent capabilities.

Overall, the results support two linked requirements of Claw-native recommendation.
A system must first ground the latent current task from fragmented and temporally mixed evidence without allowing stale context to dominate.
It must then translate that state into effective cross-source content selection and slate ranking.
Table~\ref{tab:main} establishes ClawRec's end-to-end advantage, but does not by itself isolate the contributions of unified event construction, temporal control, source extension, and marginal curation.
The detailed analyses therefore first separate these mechanisms through ablation, and then examine contextual need grounding from short prefixes, user state evolution across events, and source-role-aware retrieval under source shift.

\subsection{Detailed Analysis}

\subsubsection{Ablation Study}

We conduct an ablation study to isolate the contribution of four major mechanisms in ClawRec. All variants follow the matched setup in Section~\ref{sec:experiments}. The four variants remove unified event construction, temporal forgetting, source extension, and marginal curation, respectively.

\begin{table}[t]
\centering
\caption{Performance comparison of ClawRec and its ablated variants, each removing one major mechanism, across four state-quality dimensions and two recommendation metrics.}
\begin{tabular}{lcccccc}
\toprule
Variant & Need & Pref. & Temp. & Calib. & NDCG@20 & Hit@20 \\
\midrule
ClawRec 
& 4.57 & 3.98 & 4.29 & 3.65 & 0.6134 & 0.6944 \\

\midrule
w/o unified event construction 
& 3.86 & 3.71 & 3.68 & 3.22 & 0.5287 & 0.6112 \\

w/o temporal forgetting 
& 4.08 & 4.03 & 3.58 & 3.31 & 0.5496 & 0.6360 \\

w/o source extension 
& 4.53 & 3.92 & 4.24 & 3.65 & 0.5573 & 0.6067 \\

w/o marginal curation 
& 4.56 & 4.01 & 4.26 & 3.69 & 0.5141 & 0.6809 \\
\bottomrule
\end{tabular}
\label{tab:ablation}
\end{table}

\paragraph{Unified Events Support State Construction.}
Unified event construction has the broadest effect on state quality. Removing it reduces Need, Temp., and Calib. by 0.71, 0.61, and 0.43, respectively, indicating that structured evidence records are central to grounding current needs, maintaining temporal alignment, and controlling unsupported inference. The degradation also carries into recommendation quality: NDCG@20 decreases by 0.0847 and Hit@20 decreases by 0.0832. This pattern shows that the normalized evidence interface is a key upstream mechanism for the downstream planner.

\paragraph{Temporal Forgetting Reduces Stale-Interest Interference.}
Temporal forgetting primarily contributes to temporal alignment and current-task isolation. Removing it causes the largest Temp. drop in the table, from 4.29 to 3.58, and also lowers Need from 4.57 to 4.08. This indicates that historical needs interfere with current task grounding when temporal control is removed. The Pref. score slightly increases from 3.98 to 4.03, showing that the ablated variant retains more preference-like historical evidence. The accompanying drops in Temp., Calib., NDCG@20, and Hit@20 show the cost of that retention: historical evidence is less reliably aligned with the current recommendation context.

\paragraph{Source Extension Improves Next-Step Coverage.}
Source extension contributes mainly at the planning and retrieval stage. Removing it leaves the state metrics largely stable, while recommendation coverage changes substantially. Hit@20 drops from 0.6944 to 0.6067, the largest Hit@20 decrease among the variants, while NDCG@20 drops from 0.6134 to 0.5573. This result shows that source extension helps the system include useful next-step content under the fixed display budget, especially when the appropriate source is role-compatible but not directly observed in the current behavior prefix.

\paragraph{Marginal Curation Improves Slate Ranking.}
Marginal curation contributes mainly to slate-level ranking and organization. Removing it leaves Hit@20 close to the full model, decreasing only from 0.6944 to 0.6809, while NDCG@20 drops sharply from 0.6134 to 0.5141. This contrast shows that useful items still appear within the top-20 slate, but their ranking quality deteriorates. The state metrics remain essentially unchanged, consistent with the position of marginal curation after state construction and candidate retrieval. Its contribution is therefore concentrated in final slate construction: it prioritizes items with higher marginal value, improves complementarity, and places more useful items earlier in the rendered recommendation slate.

\begin{figure*}[t]
\centering

\begin{minipage}[t]{0.48\textwidth}
  \centering
  \includegraphics[width=\linewidth]
  {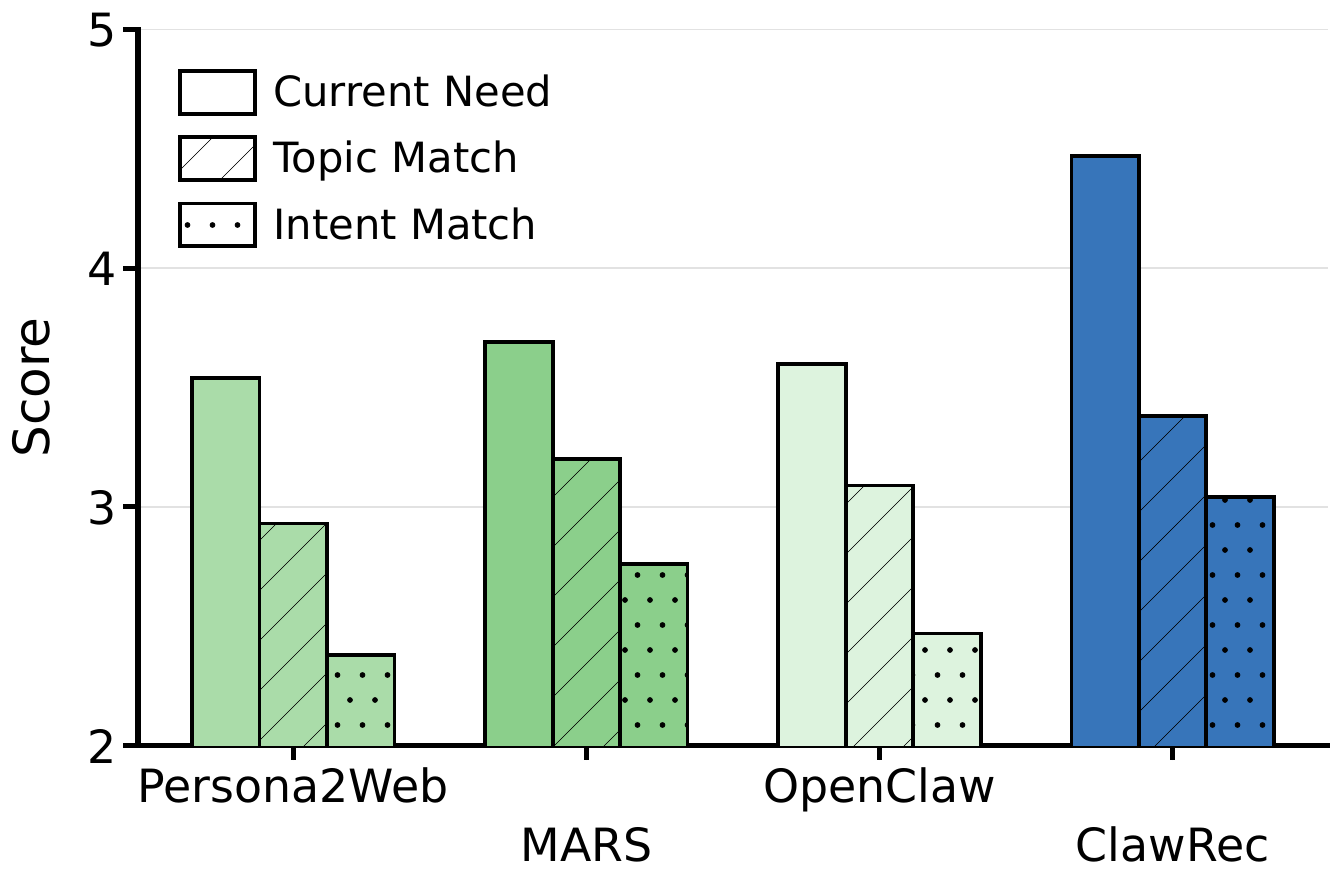}
  \caption{Current task grounding on first-event instances. ClawRec achieves the highest scores on Current Need, Topic Match, and Intent Match.}
  \label{fig:contextual_need_grounding}
\end{minipage}
\hfill
\begin{minipage}[t]{0.48\textwidth}
  \centering
  \includegraphics[width=\linewidth]
  {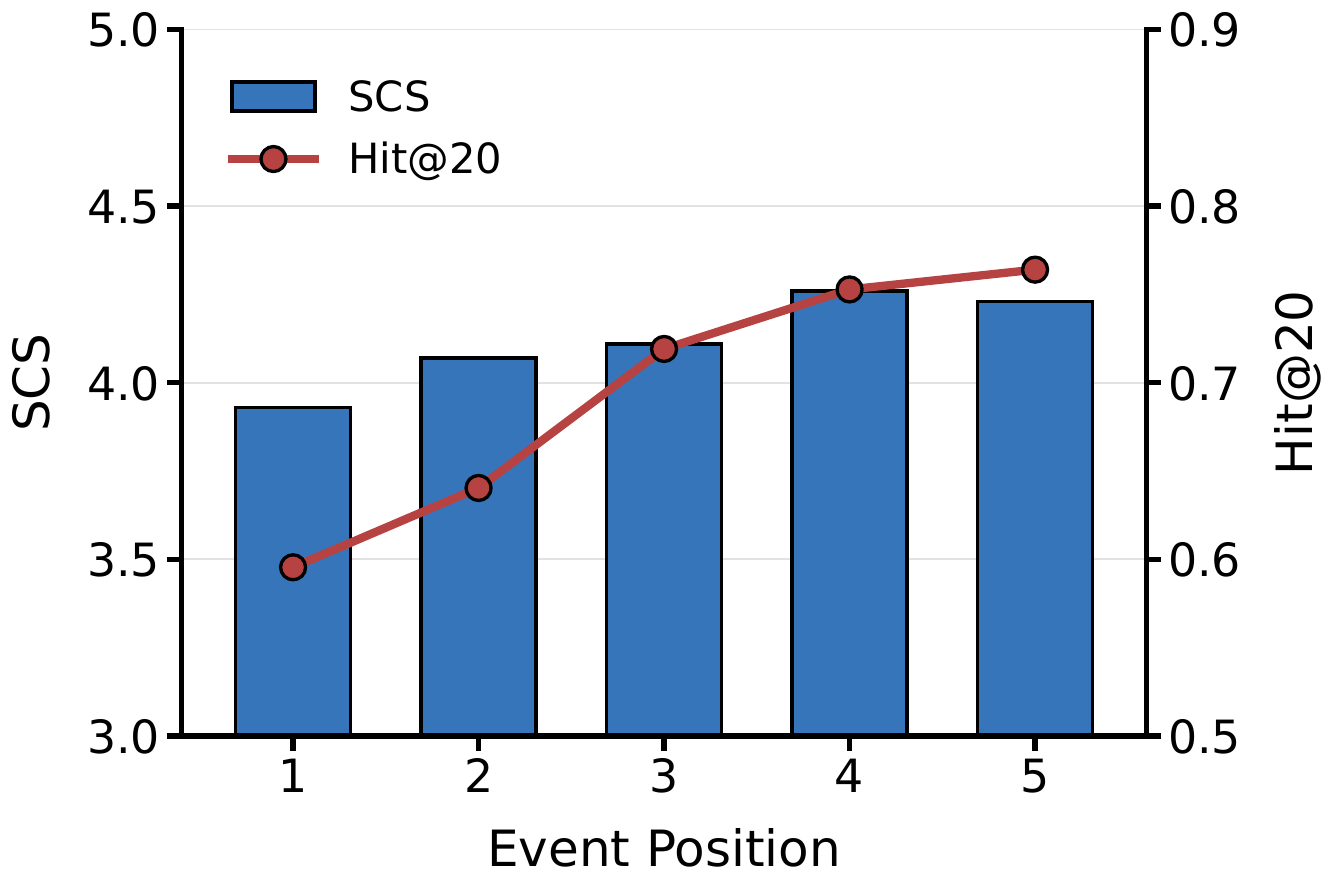}
  \caption{SCS and Hit@20 across event positions. Both metrics generally improve as cross-event evidence accumulates.}
  \label{fig:self_evolution_across_events}
\end{minipage}

\end{figure*}

\subsubsection{Current Task Grounding}

Current task grounding is most directly tested when a system cannot rely on previously accumulated cross-event history.
Personalized Web agents and memory-augmented recommendation agents commonly use user histories or persistent memories to contextualize current behavior~\cite{kim2026persona2web,shen2026agentic}.
We therefore focus on first-event instances, which represent a cold-start scenario at the cross-event level.
For these instances, \(m=1\) and \(B^{(u)}_{<1}=\varnothing\), while the system still observes the short current-event prefix \(B^{(u)}_{1,<K}\).
This setting isolates whether a method can infer the latent current task from limited within-event evidence rather than from prior life-event memory.

We compare ClawRec with agentic baselines that produce inspectable state or memory representations.
The three reported metrics form a two-level diagnostic.
Current Need evaluates whether the user state inferred from \(B^{(u)}_{1,<K}\) identifies the concrete active task.
Topic Match and Intent Match are computed on the final displayed recommended items with respect to \(h^{(u)}_1\), measuring alignment with its concrete topic and functional intent, respectively.
The latter two metrics therefore examine whether current task grounding is preserved through planning, retrieval, and curation, rather than directly evaluating the internal user state.

As shown in Figure~\ref{fig:contextual_need_grounding}, ClawRec obtains the strongest scores on all three measures, with the clearest separation appearing on Current Need.
This result indicates that its main advantage in the first-event setting lies in forming a specific task-level representation from a short heterogeneous behavior prefix.
The smaller differences on Topic Match and Intent Match suggest that the agentic baselines can often retrieve content related to the held-out topic or intent, but less reliably consolidate the available evidence into an explicit representation of the concrete active task.

This grounding profile is consistent with the staged representations in ClawRec.
Unified events preserve the topic and behavior-level intent expressed by individual actions, while user state reasoning consolidates their joint support into the active task and its constraints.
The planner then uses this task-level representation to construct retrieval directions, resulting in recommended items that remain aligned with both the concrete topic and the functional intent of the next step.

\subsubsection{Self-Evolution across Events}

Accumulated history can benefit recommendation only if supported information transfers across events without allowing completed tasks to dominate the active one.
We therefore analyze ClawRec across sequential life events.
For an instance at event position \(m\), the system observes the current-event prefix \(B^{(u)}_{m,<K}\) together with \(B^{(u)}_{<m}\), which contains \(m-1\) complete prior life-event trajectories.
Grouping instances by \(m\) provides a descriptive view of how state quality and recommendation coverage vary as more cross-event evidence becomes available.
We report SCS and Hit@20 at each event position.

Figure~\ref{fig:self_evolution_across_events} shows that both SCS and Hit@20 generally increase with event position.
Event 1 corresponds to the first-event cold-start setting analyzed above, whereas later positions provide additional evidence about supported user preferences, recurring constraints, and source-role associations.
The higher SCS at later positions indicates that ClawRec maintains a more consistent user state when richer cross-event evidence is available.
The corresponding Hit@20 trend shows that this accumulated context is also reflected in the final recommendation slate, where useful next-step content is more frequently covered within the fixed display budget.

The joint trend accords with ClawRec's self-evolution mechanism.
Repeatedly supported user preferences and constraints can be reused to contextualize later tasks, while source-role records preserve evidence about which sources have served particular functions in related contexts.
At the same time, the temporal status of state records prevents earlier completed tasks from being treated as equally active.
ClawRec therefore benefits from additional history by selectively transferring supported context rather than uniformly accumulating past behavior.

\subsubsection{Source-Role-Aware Retrieval}

Inferring the current user state is necessary but insufficient for producing an effective recommendation slate.
The active task specifies what the user is trying to accomplish, but the system must still determine which kinds of functional support are needed, which source roles can provide that support, and which concrete sources and items should be selected.
This distinction is central under source heterogeneity: different sources may be topically relevant to the same task while serving different functions, such as explanation, practical guidance, experience reference, comparison, or authoritative verification.
We therefore examine whether ClawRec translates the inferred user state into recommendations that preserve the current topic and intent while drawing on functionally appropriate sources.
The comparison covers local query continuation, lexical history expansion, generic Web-agent search, and memory-augmented agent reasoning.

\begin{figure}[t]
    \centering
    \begin{subfigure}{0.48\linewidth}
        \centering
        \includegraphics[width=\linewidth]{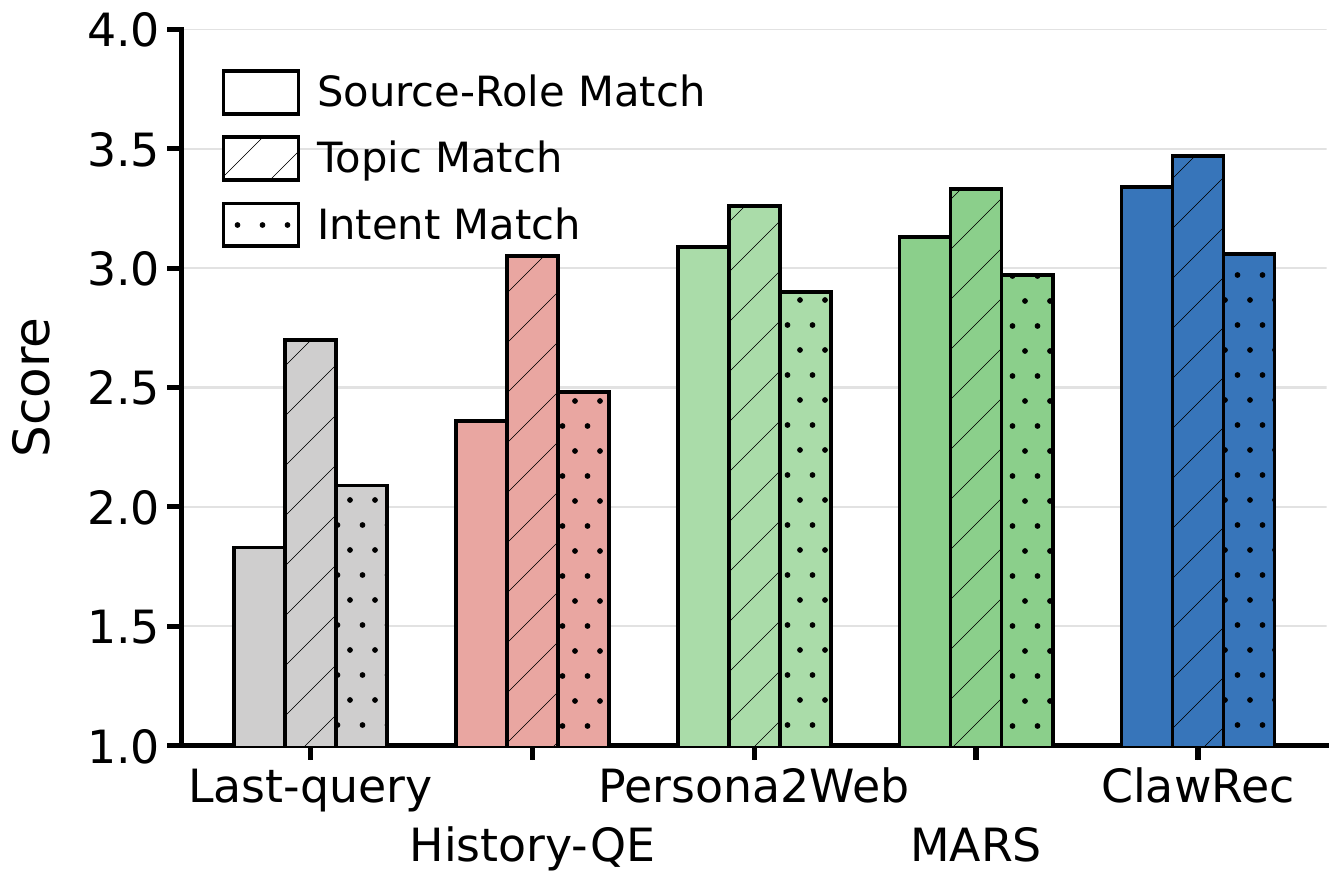}
        \caption{Source-role alignment diagnostics.}
        \label{fig:source_role_alignment}
    \end{subfigure}
    \hfill
    \begin{subfigure}{0.48\linewidth}
        \centering
        \includegraphics[width=\linewidth]{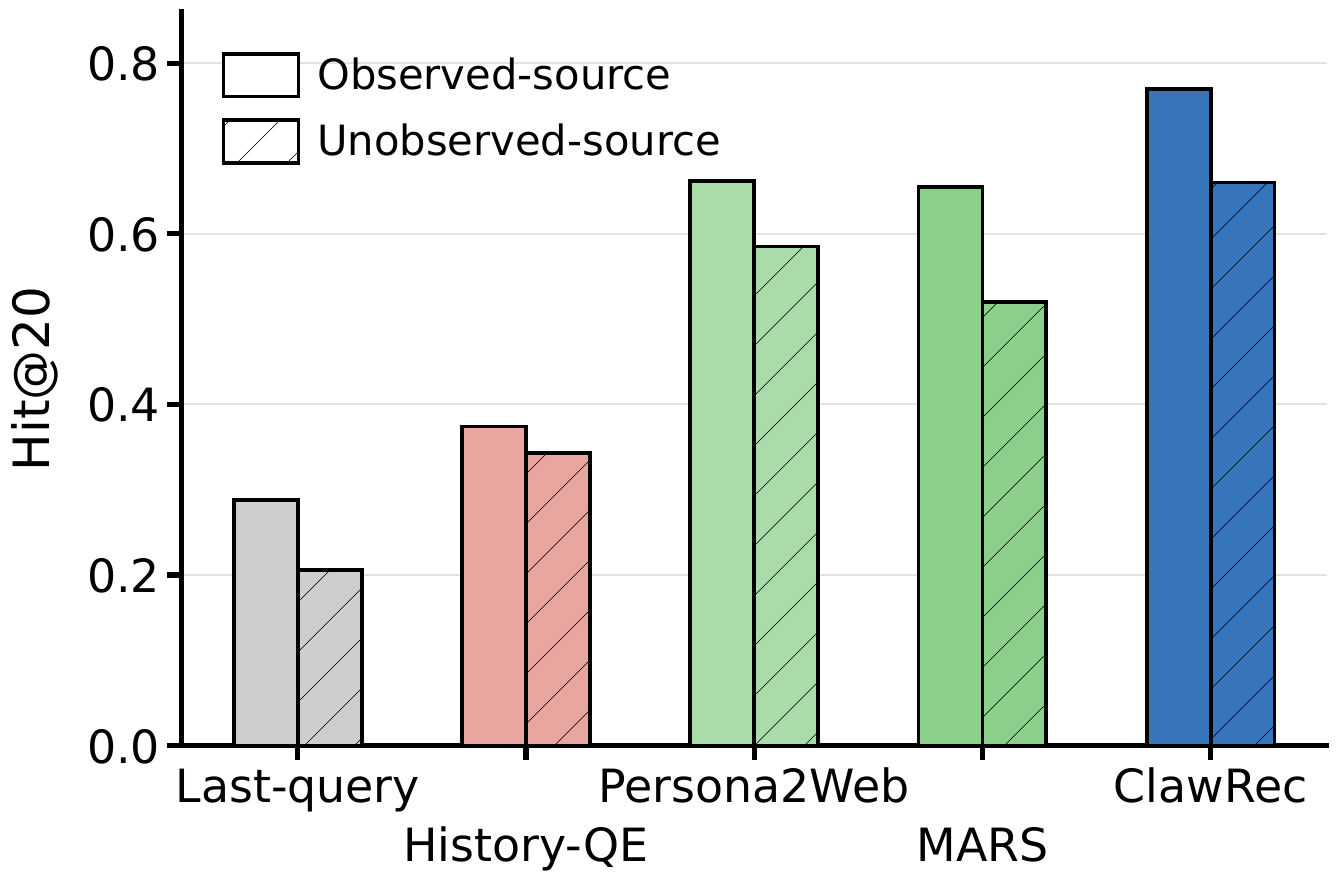}
        \caption{Hit@20 on source-shift cases.}
        \label{fig:source_shift_performance}
    \end{subfigure}
    \caption{Source-role-aware retrieval analysis. ClawRec achieves the strongest source-role alignment while maintaining comparable topic and intent alignment, and obtains the highest Hit@20 on both observed-source and unobserved-source cases.}
    \label{fig:source_role_retrieval}
\end{figure}

Figure~\ref{fig:source_role_alignment} reports Source-Role Match, Topic Match, and Intent Match on the final displayed recommended items.
Last-query Retrieval and History-QE obtain substantially lower Source-Role Match than the agentic methods, although their gaps on Topic Match and Intent Match are smaller.
This pattern suggests that recent-query continuation and lexical history aggregation can preserve topical terms or broad task intent, but are less reliable in determining what function the next source should serve.
Persona2Web and MARS narrow this gap through multi-step Web search and memory-based contextual reasoning.
ClawRec achieves the highest Source-Role Match while maintaining strong Topic Match and Intent Match, indicating that its main retrieval-side advantage lies in selecting sources whose functional roles are compatible with the active task, rather than merely retrieving more content about the same topic.

Figure~\ref{fig:source_shift_performance} further analyzes source shift by separating observed-source and unobserved-source cases.
A case is classified as observed-source when the source of \(h^{(u)}_m\) has appeared in the current-event prefix \(B^{(u)}_{m,<K}\), and as unobserved-source otherwise.
This distinction is defined only with respect to the current-event prefix; an unobserved source may still have appeared in an earlier life event.
The comparison therefore tests whether a method can move beyond source continuation from the current prefix and select another source that is functionally appropriate for the inferred task.

All methods perform worse on unobserved-source cases, confirming that source shift beyond the current prefix is a more difficult retrieval condition.
Last-query Retrieval exhibits the largest degradation because its retrieval direction is closely tied to the most recent query and its associated source context.
History-QE incorporates broader textual evidence, but does not explicitly represent the functional relationship between the current task and candidate sources.
Persona2Web and MARS perform more strongly in both groups, reflecting the benefit of multi-step search and historical context.
ClawRec achieves the highest Hit@20 on both observed-source and unobserved-source cases, with a larger advantage when the useful next-step source is absent from the current prefix.
Its role-aware planner can therefore use the inferred task to select sources beyond those exposed by the ongoing behavior chain while retaining topic and intent alignment in the final items.

\subsection{Case Study}
\label{sec:case_study}

We present a representative case to illustrate how ClawRec transforms a cross-source behavior prefix into a structured recommendation slate.
Figure~\ref{fig:case_study_trace} summarizes the execution path from observed behaviors to unified events, user state, retrieval planning, and the rendered slate.

\begin{figure}
    \centering
    \begin{tracebox}{Case Study: Cross-Source Behaviors to a Curated Slate}

\textbf{Case setting.}
{\scriptsize
ClawRec composes a recommendation slate from three cross-source behaviors visible
during inference. The subsequent Taobao behavior is withheld and shown only for
evaluation.
}

\vspace{0.7em}

\begin{minipage}[t]{0.47\linewidth}
\vspace{0pt}
\textbf{Evidence Trace: Behaviors to Active Task}

\vspace{0.35em}
{\scriptsize\ttfamily
\textbf{[1]} observe e1\par
\tracefield{source}{Zhihu}
\tracefield{action}{search}
\tracefield{content}{signs of sensory sensitivity in children}
\tracefield{intent}{understand stimulation needs}
\vspace{0.25em}

\textbf{[2]} observe e2\par
\tracefield{source}{Xiaohongshu}
\tracefield{action}{read}
\tracefield{content}{parent-child museum visits}
\tracefield{intent}{gather practical experience}
\vspace{0.25em}

\textbf{[3]} observe e3\par
\tracefield{source}{Baidu}
\tracefield{action}{search}
\tracefield{content}{museum preparation with a young child}
\tracefield{intent}{plan practical preparation}
\vspace{0.35em}

\textbf{[4]} consolidate active task\par
\tracefield{scenario}{museum visit with a young child}
\tracefield{constraint}{lower stimulation}
\tracefield{goals}{visit planning + preparation}
}
\end{minipage}
\hfill
\begin{minipage}[t]{0.49\linewidth}
\vspace{0pt}
\textbf{Decision Trace: Plan to Curated Slate}

\vspace{0.35em}
{\scriptsize\ttfamily
\textbf{[5]} plan objective o1\par
\tracefield{objective}{visit planning}
\tracefield{source role}{guidance}
\textbf{[6]} plan objective o2\par
\tracefield{objective}{coping strategies}
\tracefield{source role}{experience}
\textbf{[7]} plan objective o3\par
\tracefield{objective}{preparation options}
\tracefield{source role}{product}
\vspace{0.35em}

\textbf{[8]} retrieve candidates\par
\tracefield{guidance 1}{lower-stimulation visit guide}
\tracefield{experience}{parent-child museum experience}
\tracefield{product}{noise-reduction earmuffs}
\tracefield{guidance 2}{second quieter-hours guide}
\vspace{0.35em}

\textbf{[9]} curate candidates\par
\tracefield{retain}{experience item}
\tracefield{reason}{adds coping evidence}
\tracefield{remove}{second museum guide}
\tracefield{reason}{repeats quieter-hours advice}
\vspace{0.35em}

\textbf{[10]} render slate\par
\tracefield{guidance}{lower-stimulation visit plan}
\tracefield{experience}{parent-tested coping strategies}
\tracefield{product}{noise-reduction earmuffs}
}
\end{minipage}

\vspace{0.6em}
\tcbline
\vspace{0.4em}

\textbf{Held-out Evaluation (Target Hidden during Inference)}

\vspace{0.2em}
{\scriptsize\ttfamily
\textbf{[11]} compare slate with held-out target\par
\tracefield{target source}{Taobao}
\tracefield{target action}{search}
\tracefield{target content}{children's noise-reduction earmuffs and calming items}
\tracefield{product match}{noise-reduction earmuffs}
\tracefield{task support}{guidance + experience complement the same active task}
}

\end{tracebox}
    \caption{Case study of ClawRec's recommendation process, from cross-source behavioral evidence and active-task inference to role-aware planning, marginal curation, and held-out evaluation.}
    \label{fig:case_study_trace}
\end{figure}

\paragraph{Case Overview.}
The observed behavior prefix contains three related behaviors from different sources.
The user first searches Zhihu for signs of sensory sensitivity in children, then reads a Xiaohongshu post about parent--child museum visits, and finally searches Baidu for museum preparation with a young child.
The first behavior indicates a concern about stimulation needs, the second places that concern in a museum-visit scenario, and the third expresses a practical preparation objective.
Together, they support the current task of planning a lower-stimulation museum visit with a young child and identifying practical preparation steps.
Under the leave-one-out protocol, the subsequent Taobao search for children's noise-reduction earmuffs and calming items is hidden during inference and used as the held-out target.

\paragraph{Execution Trace.}

As shown in Figure~\ref{fig:case_study_trace}, the three source-specific behaviors are represented as separate unified events with their respective topics and behavior-level intents.
User state reasoning consolidates their joint evidence into an active task with the museum-visit context and lower-stimulation constraint.
The planner then instantiates support objectives for visit planning, practical experiences, and preparation options, paired with the guidance, experience, and product source roles.
During marginal curation, the experience item is retained because it contributes coping evidence, whereas a second guide is removed because it repeats the quieter-hours advice.
The rendered slate consequently combines planning guidance, experience-based strategies, and a concrete preparation aid.

\paragraph{Analysis.}

The case shows how ClawRec uses intermediate representations to connect heterogeneous evidence with complementary recommendations.
Unified events preserve the local semantics and provenance of each behavior, while the active task captures their shared scenario and constraint.
Planning operates on this task-level representation rather than on the latest query alone, allowing the final slate to cover distinct functional needs.
The earmuff item is closely aligned with the held-out next step, while the other items provide complementary support for the same current task.

\section{Related Work}

\paragraph{Cross-Domain Recommendation.}
Cross-domain recommendation transfers knowledge across related domains to alleviate sparsity and cold-start problems, with settings spanning single-target, dual-target, and multi-domain recommendation~\cite{zang2022survey}.
Early approaches transferred cluster-level rating patterns without requiring overlapping users or items, or pooled multiple rating matrices through shared latent structures~\cite{li2009can,li2009transfer}, while coordinate-system transfer accommodated heterogeneous feedback such as ratings and clicks~\cite{pan2010transfer}.
Later work jointly modeled browsing and search signals with items across services, or learned cross-domain connections, latent-factor mappings, and personalized transfer functions for overlapping users and cold-start targets~\cite{elkahky2015multiview,man2017crossdomain,zhu2018deep,hu2018conet,zhu2022personalized}.
Auxiliary review aspects and interactions from multiple social networks further enriched target-domain user modeling~\cite{zhao2020catn,perera2018lstm}.
Multi-domain recommenders instead balance shared knowledge with domain-specific distributions and optimization conflicts across predefined business scenarios~\cite{sheng2021one,luo2023mamdr}.
These lines broaden the evidence available to recommendation, but typically assume predefined domains and optimize recommendation within one or more domain-specific spaces through shared structures, aligned entities, or learned mappings.
ClawRec addresses a different setting: it infers the temporally active task from heterogeneous cross-platform behaviors and organizes next-step recommendations from agent-accessible, functionally heterogeneous sources.

\paragraph{LLM-Based Recommender Systems.}
LLMs are increasingly explored in recommendation tasks due to their strong semantic understanding and reasoning abilities~\cite{wu2024survey,liu2024large}.
Early studies reformulate recommendation as text prompting and directly apply LLMs to recommendation. These methods usually take user context as input and prompt LLMs to infer user preferences or generate recommendation results~\cite{hou2024large,geng2022recommendation,gao2023chat,zhang2026recommendation}.
Directly prompting LLMs, however, does not fully bridge the gap between general language modeling and domain-specific recommendation behavior. To address this issue, subsequent studies explore several adaptation strategies, including incorporating LLM knowledge into recommendation models~\cite{xi2024towards,zheng2024adapting} and fine-tuning LLMs on recommendation data~\cite{bao2023tallrec,chen2024softmax}.
Recently, LLM-based agents are introduced into recommender systems~\cite{peng2025survey,zhu2025recommender,zhang2025survey}. Some studies use agents to perform or assist recommendation tasks~\cite{wu2025starec,huang2025mr,cai2026mgfrec,ou2026deep}. Other studies use agents to simulate recommendation environments, providing dynamic interaction processes for collaborative filtering and the study of recommendation behavior~\cite{zhang2024agentcf,liu2025agentcf++,xia2026multi}.
However, these methods usually operate in bounded, platform-centric settings, whereas ClawRec studies task-level needs that must be inferred before recommendation.

\paragraph{Personalized Web Agents.}
Autonomous agents for Web and personal applications aim to help users continuously acquire information and execute tasks in complex digital environments~\cite{xie2024osworld,yao2022webshop,gou2026mind2web}.
Research on Web agents establishes a foundation for enabling agents to understand instructions, operate interfaces, and complete multi-step tasks in Web environments~\cite{zhou2023webarena,deng2023mind2web,lu2024weblinx,he2024webvoyager}. However, its typical setting usually starts from a given user goal and gives limited attention to how the goal itself emerges, changes, and becomes identifiable from users' everyday information behavior.
Research on personal agents extends agents from one-shot task execution to personal workspaces with long-term memory~\cite{zhong2024memorybank,wu2024longmemeval,xu2026mem,chhikara2025mem0} and persistent environmental state~\cite{openclaw2026}. In this paradigm, user context accumulates across tasks and shapes subsequent actions, which provides a new modeling perspective for recommender systems.
Personalized Web agents further show that Web task execution can be adapted according to user profiles~\cite{salemi2024lamp,kumar2024longlamp,chen2024apollonion,zhao2025llms}, historical preferences~\cite{kim2026persona2web,cai2025large}, or interaction feedback~\cite{liang2026learning}, thereby producing actions or results that better fit the user's personal context in the open Web.
In comparison, ClawRec extends this class of agentic systems to recommendation scenarios where the central objective is to identify and serve evolving information needs, rather than only execute a pre-specified task.

\section{Conclusion}

This paper presented ClawRec, a Claw-native recommender system that extends recommendation beyond platform-local boundaries to support evolving user needs in an agent-accessible digital environment.
ClawRec makes this broader environment actionable for recommendation by turning fragmented and temporally mixed behavioral evidence into an evolving understanding of user needs, then using that understanding to organize complementary support from functionally heterogeneous sources.
For rigorous evaluation, we introduced ClawRec-SimBench, which organizes each simulated user's history as a sequence of concrete life events.
Every event defines an immediate information-seeking or decision-support task and is realized as a cross-platform behavior trajectory.
The benchmark tests current-task inference, cross-event state evolution, temporal handling of stale context, and recommendation of content that supports a held-out next behavior.
Experiments show that ClawRec improves both user state quality and recommendation quality over non-LLM, direct LLM, and agentic baselines, with ablations confirming the roles of unified event construction, temporal forgetting, source extension, and marginal curation.
The current study is bounded by the scope of ClawRec-SimBench, where user trajectories are generated from simulated life events and evaluated through a leave-one-out protocol under predefined source, budget, and judging settings, so the results primarily characterize task-level next-step support rather than long-term real-world deployment effects.
Future work should extend the evaluation to longitudinal user-facing studies, broader multilingual and multimodal source environments, and richer feedback loops, while preserving the evidence grounding, privacy filtering, and auditability required for practical deployment.


\bibliographystyle{unsrt}
\bibliography{ref,manual_ref}

\appendix
\section{Privacy and Audit Controls}
\label{app:controls}

\paragraph{Privacy Controls.}
ClawRec uses local-first processing for behavior evidence.
Raw browser evidence is extracted only from user-authorized local sources.
Sensitive surfaces, such as communication, payment, order, account, private-dashboard, and authentication pages, are filtered before semantic inference.
Entity redaction replaces names, phone numbers, addresses, account identifiers, and other detected sensitive strings with consistent local placeholders before LLM calls.
The user state stores compact topics, confidence, status, and evidence handles rather than full browsing logs or credentials.
Privacy filtering is applied before unified event construction, during state-patch validation, and again before candidate display.

\paragraph{Audit Trail.}
Each rendered recommendation slate \(\mathcal{R}_t\) links to an audit trail containing the unified events, state slots, retrieval tasks, executed queries, candidate rationales, curation decisions, and feedback records that produced it.
Users can configure observable sources, time windows, and denied domains.
They can delete event ranges, edit or remove state slots, block topics, and disable triggers.
These mechanisms provide practical transparency and control, but they are not formal privacy guarantees such as differential privacy or cryptographic anonymization.

\section{Evaluation Prompts}
\label{app:evaluation_prompts}

\begin{PromptBox}{State Quality Prompt}
\begin{lstlisting}
# System prompt:

You are evaluating the quality of a candidate User State. Judge only whether the candidate is faithful to the observed behavior available at the evaluation horizon. Do not use the held-out behavior or any future event.

Assign scores for Current Need, Preference, Temporal Alignment, and Calibration. Do not compute an aggregate score.

# User prompt:

You are grading a candidate User State against the observed behavior prefix.

The observed behavior prefix contains the behavioral evidence available to the system at the evaluation horizon. The candidate may use different wording, but its claims must be semantically supported by this evidence.

## Current Need

Evaluate whether the candidate correctly identifies the user's current needs and active goals.

5: The active needs are precise, specific, actionable, and consistent with the observed behavior. Important situations, constraints, and goals are covered.
4: The active needs are basically correct and actionable, with only minor constraints or sub-goals missing.
3: The broad intent is correct, but the specific target is vague, incomplete, or partly shifted.
2: Only the broad domain is captured, without the concrete situation or immediate goal.
1: The identified needs are wrong, contradictory, unrelated, or dominated by irrelevant needs.

## Preference

Evaluate whether the candidate correctly represents longer-running preferences, constraints, recurring patterns, and platform or source tendencies.

5: Important preferences and constraints supported by the observed behavior are captured accurately. Stable preferences are distinguished from short-lived needs.
4: Preferences are mostly correct, with minor omissions or slight overgeneralization.
3: Some valid preferences are present, but important constraints or recurring patterns are missing or generic.
2: The candidate mainly states broad categories or habits without sufficient user-specific support.
1: The preferences are unsupported, fabricated, inferred from weak evidence, or contradicted by the observed behavior.

## Temporal Alignment

Evaluate whether the candidate correctly distinguishes active needs, cooling interests, expired interests, suppressed interests, and completed one-off tasks at the evaluation horizon.

5: Temporal status is faithful to the observed sequence. Current needs remain current, while older, completed, cooling, or suppressed interests are represented at the appropriate strength.
4: Temporal handling is mostly correct, with one minor stale-versus-active mismatch.
3: The rough sequence is recognized, but several active and historical states are blurred.
2: Temporal distinctions are weak. Older or completed goals are frequently treated as current, or current needs are diluted by history.
1: Temporal alignment is absent or mostly incorrect.

## Calibration

Evaluate whether the candidate avoids unsupported inference and overconfidence.

5: All or nearly all claims are supported by the observed behavior and appropriately qualified. Weak or one-off signals are not promoted into durable conclusions.
4: Claims are mostly well supported, with one or two mild overstatements.
3: Several claims are supported, but some are overly confident, overly durable, or weakly grounded.
2: Unsupported inference or overgeneralization occurs frequently.
1: Most claims are speculative, fabricated, contradictory, or presented with unjustified certainty.

## Evaluation Rules

Do not reward plausible but unsupported details.
Do not penalize wording differences when the meaning is faithful.
Penalize missing major current needs, treating one-off behavior as a durable preference, confusing historical and current interests, and making unsupported demographic, medical, identity, or lifestyle claims.

## Output

Output one integer score from 1 to 5 for each of the following dimensions: Current Need, Preference, Temporal Alignment, Calibration.

## Observed Behavior Prefix
{{OBSERVED_BEHAVIOR_PREFIX}}

## Candidate User State
{{CANDIDATE_USER_STATE}}
\end{lstlisting}
\end{PromptBox}

\begin{PromptBox}{Recommendation Relevance Prompt}
\begin{lstlisting}
# System prompt:

You are a strict trajectory recommendation judge. Evaluate whether each final displayed recommendation card can semantically substitute for or directly support the user's held-out next-step behavior.

Use only the observed behavior prefix, the held-out behavior, and the displayed recommendation cards. Output only the relevance score rel for each card.

# User prompt:

You are grading final displayed recommendation cards against a held-out next-step behavior.

The goal is not to predict the exact URL, page, or item. A card may receive a high score when it supports substantially the same next-step need, even if its wording or specific item differs from the held-out behavior.

## Recommendation Relevance

Assign rel as an integer from 1 to 5 for each card.

5: Near-perfect semantic substitute. The card would satisfy almost the same next-step need as the held-out behavior, matching the topic, intent, useful action, and source role.
4: Strong substitute. The card is useful for the same task and mostly matches the held-out behavior, with only minor specificity or source-role gaps.
3: Plausible partial substitute. The card is related to the same event need but misses an important constraint, intent, source role, or actionable component.
2: Weak partial relation. The card concerns the same broad area but is only indirectly useful or too vague for the held-out next step.
1: The card is unrelated, misleading, stale, not useful, or only broadly related without supporting the specific held-out behavior.

## Evaluation Rules

Judge each card as it is finally displayed.

Do not require an exact URL or item match.

Reward semantic substitution in topic, intent, source role, and task usefulness.

Penalize generic cards that are only loosely related to the held-out behavior.

Penalize cards that continue stale or completed needs instead of supporting the held-out next step.

## Output

Output only rel for each recommendation card.

## Observed Behavior Prefix
{{OBSERVED_BEHAVIOR_PREFIX}}

## Held-Out Behavior
{{HELDOUT_BEHAVIOR}}

## Final Displayed Recommendation Cards
{{FINAL_RECOMMENDATION_CARDS}}
\end{lstlisting}
\end{PromptBox}

\begin{PromptBox}{Diagnostic Metric Prompt}
\begin{lstlisting}
# System prompt:

You are evaluating the diagnostic alignment of final displayed recommendation cards with a held-out next-step behavior.

Evaluate only Source-Role Match, Topic Match, and Intent Match. Judge the final recommendation cards themselves, not a content plan, Search Instruction, or hypothetical search result.

# User prompt:

You are grading final displayed recommendation cards against a held-out next-step behavior.

Use only the observed behavior prefix, the held-out behavior, and the cards as finally displayed.

## Source-Role Match

Evaluate whether the card comes from a source that plays an appropriate functional role for the held-out next step, such as official verification, explanation, tutorial, experience reference, product comparison, or transaction.

5: The source has an ideal role or platform for the held-out next step.
4: The source role is compatible with the held-out need, with only a minor affordance gap.
3: The source role is acceptable but not ideal for the held-out next step.
2: The source provides weakly useful support for an adjacent need.
1: The source role is wrong or unhelpful for the held-out need.

## Topic Match

Evaluate whether the card matches the concrete topic of the held-out behavior rather than only sharing a broad domain.

5: The card has a near-exact match with the held-out topic.
4: The card matches the specific target topic but misses a minor constraint.
3: The card covers a related subtopic but not the specific held-out need.
2: The card shares only broad domain overlap.
1: The card concerns the wrong topic.

## Intent Match

Evaluate whether the card supports the same user intent as the held-out behavior, such as explanation, comparison, purchase planning, tutorial, troubleshooting, experience seeking, or official verification.

5: The card has a near-exact match with the held-out task intent.
4: The card supports the same task intent but has minor ambiguity or misses a minor constraint.
3: The card partially supports the same task but has an important intent gap.
2: The card supports only an adjacent or background intent.
1: The card supports the wrong intent.

## Evaluation Rules

Judge the recommendation card itself rather than what a search instruction might retrieve.

Do not require an exact URL or item match.

Evaluate Source-Role Match, Topic Match, and Intent Match independently.

## Output

For each recommendation card, output only:

- source_role_match
- topic_match
- intent_match

Each score must be an integer from 1 to 5.

## Observed Behavior Prefix
{{OBSERVED_BEHAVIOR_PREFIX}}

## Held-Out Behavior
{{HELDOUT_BEHAVIOR}}

## Final Displayed Recommendation Cards
{{FINAL_RECOMMENDATION_CARDS}}
\end{lstlisting}
\end{PromptBox}

\end{document}